\documentclass[trackchanges, twocolumn]{aastex7}

\usepackage[T1]{fontenc}
\usepackage[utf8]{inputenc} 
\usepackage{float}

\begin{document}

\title{\emph{JWST} and Keck Observations of the Off-Nuclear TDE AT\,2024tvd: A Massive Nuclear Star Cluster and Minor-Merger Origin for its Black Hole}

\author[orcid=0000-0002-1092-6806,sname='Patra']{Kishore C. Patra}
\affiliation{Department of Astronomy and Astrophysics, UC Santa Cruz, 1156 High Street, Santa Cruz, CA 95064 }
\affiliation{University of California Observatories, 1156 High Street, Santa Cruz, CA 95064}
\email[show]{kcpatra@ucsc.edu}  

\author[orcid=0000-0002-2445-5275, sname='Foley']{Ryan J. Foley} 
\affiliation{Department of Astronomy and Astrophysics, UC Santa Cruz, 1156 High Street, Santa Cruz, CA 95064 }
\email{foley@ucsc.edu}

\author[orcid=0000-0003-1714-7415]{Nicholas Earl} 
\affiliation{Department of Astronomy, University of Illinois at Urbana-Champaign, 1002 W. Green St., IL 61801, USA}
\email{nmearl2@illinois.edu}

\author[orcid=0000-0002-5680-4660]{Kyle~W.~Davis} 
\affiliation{Department of Astronomy and Astrophysics, UC Santa Cruz, 1156 High Street, Santa Cruz, CA 95064 }
\email{kywdavis@ucsc.edu}

\author[orcid=0000-0003-2558-3102]{Enrico~Ramirez-Ruiz} 
\affiliation{Department of Astronomy and Astrophysics, UC Santa Cruz, 1156 High Street, Santa Cruz, CA 95064 }
\email{enrico@ucolick.org}

\author[0000-0002-5814-4061]{V.~Ashley~Villar}
\affiliation{The NSF AI Institute for Artificial Intelligence and Fundamental Interactions}
\affiliation{Center for Astrophysics \textbar{} Harvard \& Smithsonian, 60 Garden Street Cambridge, MA 02138 \\}
\email{ashleyvillar@cfa.harvard.edu}

\author[orcid=0000-0001-6395-6702]{Sebastian Gomez} 
\affiliation{Center for Astrophysics \textbar{} Harvard \& Smithsonian, 60 Garden Street Cambridge, MA 02138 \\}
\affiliation{University of Texas Austin, 2515 Speedway, Stop C1400
Austin, TX 78712, USA}
\email{sebastian.gomez@austin.utexas.edu}

\author[0000-0002-4235-7337]{K. Decker French} 
\affiliation{Department of Astronomy, University of Illinois at Urbana-Champaign, 1002 W. Green St., IL 61801, USA}
\email{deckerkf@illinois.edu}

\author[orcid=0000-0002-5748-4558]{Kirsty Taggart} 
\affiliation{Department of Astronomy and Astrophysics, UC Santa Cruz, 1156 High Street, Santa Cruz, CA 95064 }
\email{k.taggart@ucsc.edu}

\author[orcid=0000-0002-6688-3307]{Prasiddha Arunachalam} 
\affiliation{Department of Astronomy and Astrophysics, UC Santa Cruz, 1156 High Street, Santa Cruz, CA 95064 }
\email{parunach@ucsc.edu}

\author[orcid=0000-0002-9946-4635]{Phillip Macias} 
\affiliation{Department of Astronomy and Astrophysics, UC Santa Cruz, 1156 High Street, Santa Cruz, CA 95064 }
\email{pmacias@ucsc.edu}

\author[orcid=0009-0005-1871-7856]{Ravjit Kaur} 
\affiliation{Department of Astronomy and Astrophysics, UC Santa Cruz, 1156 High Street, Santa Cruz, CA 95064 }
\email{rkaur44@ucsc.edu}

\author[orcid=0000-0002-1481-4676]{Samaporn Tinyanont} 
\affiliation{National Astronomical Research Institute of Thailand, 260 Moo 4, Donkaew, Maerim, Chiang Mai, 50180, Thailand}
\email{samaporn@narit.or.th}

%\collaboration{all}{The Terra Mater collaboration}

\begin{abstract}
We present \textit{JWST}/NIRSpec and NIRCam observations of the first optically selected off-nuclear tidal disruption event (TDE), AT\,2024tvd, along with Keck/KCWI integral field unit spectroscopy. The spectra show broad H and He emission lines that are characteristic of a TDE. Stellar kinematics show smooth host-galaxy morphology and ordered bulge rotation, with no evidence of disturbances in velocity, dispersion, age or metallicity space. We construct the first quasi-simultaneous spectral-energy distribution (SED) from X-rays to infrared for a TDE and decompose it into three components: the TDE accretion flow, an unresolved nuclear star cluster (NSC), and heated dust emission. The accretion component implies a black hole mass of $\log(M_\bullet/M_\odot) = 5.50\pm 0.04$, an instantaneous super-Eddington accretion rate of $\log (\dot{M}/M_{\odot} ~\rm yr^{-1}) = -1.22 \pm 0.04$, and an outer disk photosphere radius of $\log(r_{\rm out}/r_{g}) = 3.8 \pm 0.1$. The dust emission is well described by a blackbody with $T_{\rm dust} = 873\pm 15$~K and peak luminosity $\log (L_{\rm dust}/\rm erg\,s^{-1}) = 40.80\pm 0.01$, consistent with a dust echo near the sublimation radius. The SED is best fit when including additional stellar emission above the galaxy background at the TDE location, corresponding to $\log(M_{\star}/M_\odot) = 7.97^{+0.16}_{-0.26}$, which we interpret as a massive NSC or an ultra-compact dwarf galaxy. 
These results support a minor–merger origin for the MBH responsible for the TDE over scenarios involving gravitational recoil or dynamical ejection from the nucleus.
\end{abstract}

% \keywords{\uat{Galaxies}{573} --- \uat{Cosmology}{343} --- \uat{High Energy astrophysics}{739} --- \uat{Interstellar medium}{847} --- \uat{Stellar astronomy}{1583} --- \uat{Solar physics}{1476}}


\section{Introduction} 

Tidal disruption events (TDEs) occur when a star in a near-parabolic orbit gets close enough to a massive black hole (MBH) to cross its tidal radius, $R_{\rm T} \approx R_{\star} (M_{\bullet}/M_{\star})^{1/3}$, where $M_{\bullet}$ is the mass of the MBH, and $R_{\star}$ and $M_{\star}$ are the radius and the mass of the disrupted star, respectively \citep{Hills_1975}. Within this radius, the tidal forces exerted by the MBH exceed the star's self-gravity, causing the star to be disrupted \citep{Guillochon_Ramirez_2013}. The stellar debris, after disruption, acquires a distribution of specific orbital energies due to the gradient of the MBH’s gravitational potential across the star at the tidal radius. Consequently, approximately half of the debris becomes gravitationally bound to the MBH and gradually returns, forming an accretion disk \citep{Rees_1988}. The remaining half escapes as unbound material. The infalling matter fuels the accretion disk, generating a luminous transient flare observable across much of the electromagnetic spectrum, including X-ray, ultraviolet (UV), optical, and radio wavelengths (see for e.g., \citealt{Gezari_2021_tderev, Yao_Yuhan_2023_TDEs, Hammerstein_2023_ZTFTDEs, Guillochon_etal_2014, 2017_Auchettl}).

TDE flares are powerful signposts of otherwise dormant MBHs, particularly at masses below $10^{8}\,M_{\odot}$---they provide a unique means to probe MBH demographics and constrain fundamental properties such as mass and possibly spin in systems that would otherwise remain undetected (e.g., \citealt{2012_Macleod, Stone_2016_TDE_MBH, Komossa_2015_TDErev, Rees_1988, vanvelzen21_TDE_tdes, Mummery_2024_BH-TDE_scaling, 2019_Mockler}).  Up until recently, nearly all confirmed TDEs were discovered in the nuclei of galaxies, where MBHs typically reside. However, galaxy-merger simulations predict a substantial population of MBHs that are offset from galactic centers following mergers \citep{Ricarte_2021_wandBHA, Ricarte_2021_wandBHB, 2023_Mockler, 2024_Melchor}. TDEs produced by such ``wandering'' MBHs offer a unique opportunity to study accretion disks in relatively clean environments, away from the heavy stellar contamination of the bright galactic nuclei. These events also provide crucial insight into how MBHs migrate, merge, and grow over cosmic time (see e.g., \citealt{Greene_2020_IMBHrev} for review). Two off-nuclear TDE \textit{candidates} have previously been identified in X-ray surveys: 3XMM J215022.4$-$055108 \citep{Lin_2018_J215, Lin_2020_J215} and EP240222a \citep{Jin_2025_EP24}. Recently, AT\,2024tvd became the first bona fide \textit{optically selected} off-nuclear TDE \citep{Yao_Yuhan_2025_24tvd}. 

Initially discovered by the Zwicky Transient Facility (ZTF; \citealp{Bellm_etal_2019}) and spectroscopically classified as a TDE by \citet{Faris_2024_24tvdTNS}, high-resolution \textit{HST} imaging revealed that AT\,2024tvd is offset by $0.91''$ ($0.81$~kpc) from the host galaxy nucleus. This result was independently confirmed by \textit{Chandra} and VLA detections of the X-ray and radio counterparts \citep{Yao_Yuhan_2025_24tvd}. The transient shows all hallmark signatures of a TDE: a hot UV/optical blackbody with $L_{\rm bb}\approx 6\times10^{43}$~erg~s$^{-1}$, broad Balmer emission lines, and luminous, variable soft X-ray emission ($L_{X,{\rm peak}}\approx 3\times10^{43}$~erg~s$^{-1}$). The inferred black hole mass of $\sim10^{6}M_\odot$ is at least an order of magnitude below the central black hole mass of the host galaxy ($\gtrsim10^{8}M_\odot$), estimated from the $M_{\bullet}-\sigma$ relation \citep{Greene_2020_IMBHrev}. 
By comparing AT\,2024tvd with the two previously known X-ray–selected off-nuclear TDEs, \citet{Yao_Yuhan_2025_24tvd} argue that such events preferentially occur in massive galaxies ($M_\star \approx 10^{10.9}M_\odot$), consistent with cosmological predictions that the abundance of offset black holes scales with host halo mass \citep{Ricarte_2021_wandBHB}.

Subsequently, broadband radio and millimeter monitoring of AT\,2024tvd by \citet{Sfaradi_2025_24tvdradio} established it as a radio-bright off-nuclear TDE. AT\,2024tvd exhibits two distinct, double-peaked radio flares with remarkably fast temporal evolution---among the fastest observed in any TDE \citep{2019_Mockler}. Similar double-peaked radio flares have also been observed, albeit on longer timescales, in some nuclear TDEs such as AT\,2020zso and ASASSN-15oi \citep{2025arXiv250914317C, 2025ApJ...983...29H}. \citet{Sfaradi_2025_24tvdradio} argue that these radio flares cannot be easily explained by a single prompt outflow. Instead, the data are consistent with either a single delayed outflow launched $\sim 80$ days after disruption, or two separate outflows with the second beginning at $\sim 170$--190 days.

A key question raised by AT\,2024tvd is the origin of its offset MBH. One possibility is that the MBH was displaced from the nucleus through gravitational-wave recoil following a black hole merger or via dynamical ejection during three-body interactions (for e.g., \citealt{2016MNRAS.456..961B, Komossa_2012_recoilBHrev, Volonteri_Perna_2005_ejectMBH}). 
Alternatively, the offset MBH could be the product of a minor galaxy merger event, currently in the dynamical friction phase of its inspiral toward the host nucleus (e.g., \citealt{Tremmel_2018_wandMBH}). Distinguishing between these scenarios may require investigation of the stellar kinematics around the TDE, which could reveal signatures of MBH motion with respect to the stellar field and potentially signatures of past or current galaxy mergers in velocity space.  

Another important question concerns the stellar reservoir that fuels such an offset TDE. In particular, the presence of a dense nuclear star cluster (NSC) around the MBH could enhance the disruption rate by increasing the local stellar phase density and providing efficient channels for loss-cone refilling (e.g., \citealt{Magorrian_1999_TDE-NSC, Wang_merritt_2004_TDE-NSC, 2009_Ramirez-Ruiz, Stone_2016_TDE_MBH, 2016_MacLeod, 2020_Pfister, 2024_Polkas, 2024_Hannah, 2025_Hannah,Rozner2025}). One path for detecting an unresolved NSC is through multi-wavelength spectral energy distribution (SED) analysis, with particular emphasis on the infrared region, where stellar emission can dominate over the TDE accretion component. However, such studies have been challenging for nuclear TDEs owing to overwhelming infrared contamination from stars in the host nucleus.  

In this work, we focus on the two aforementioned questions: the origin of the off-nuclear MBH and the presence of an NSC in AT\,2024tvd. Using integral-field spectroscopy with \textit{JWST}/NIRSpec and Keck/KCWI, we study the stellar kinematics of the host and search for signatures of MBH displacement. We also construct and analyze a quasi-simultaneous SED spanning the X-ray to the infrared, with the goal of identifying possible NSC contribution and constraining TDE accretion properties. 
The paper is organized as follows. In Section \ref{sec:obs_and_data_red} we present the observations and data reduction. Section \ref{sec:analysis_results} describes the analysis and results. In Section \ref{sec:discussion} we discuss the results, and 
finally, in Section \ref{sec:conclusion} we summarize our conclusions.

\section{Observations and Data Reduction}
\label{sec:obs_and_data_red}

AT\,2024tvd was observed with \textit{JWST}’s NIRCam and NIRSpec Integral Field Unit (IFU) as part of a Cycle 3 Director's Discretionary (DD) program (ID: 9249; PI: Kishore Patra). The observations were carried out on 2025 April 03 (UT; MJD 60768), which is $+199$ days relative to the $g-$band peak \citep{Yao_Yuhan_2025_24tvd}.

\subsection{\textit{JWST} NIRCam}
We obtained near- to mid-infrared imaging with NIRCam to search for subtle signs of disturbance in the host galaxy---such as faint tidal features, stellar streams, or trails that could hint at a recent merger or the motion of the massive black hole through the stellar field. Observations were taken simultaneously in four filters (F090W, F150W, F277W, and F444W) using the short- (SW) and long-wavelength (LW) channels. The data were acquired with the \texttt{BRIGHT2} readout pattern and a 4-point primary dither combined with a 3-point subpixel dither sequence, designed to improve point-spread function (PSF) sampling and achieve an effective spatial resolution of $\sim0.05\arcsec$. Each exposure used four groups per integration and one integration per exposure, yielding a total on-source time of 1031~s across 12 dithered frames.

The NIRCam data were processed using the standard \textit{JWST} Calibration Pipeline (Version~1.17.1; CRDS context \texttt{jwst\_1322.pmap}; \citealp{Bushouse_2023_pipeline}). In Stage~1, we applied the usual detector-level calibrations---reference pixel, bias, dark current, and non-linearity corrections---followed by the conversion of the up-the-ramp samples into count-rate images. Because NIRCam detectors are known to exhibit noticeable 1/$f$ noise, we enabled the \texttt{clean\_flicker\_noise} option to mitigate this effect. We experimented with different settings of the \texttt{fit\_method}, \texttt{background\_method}, and \texttt{fit\_by\_channel} parameters to assess how they influence the results. The cleanest output was obtained with both \texttt{fit\_method} and \texttt{background\_method} set to \texttt{median} and \texttt{fit\_by\_channel} set to \texttt{False}.

Stage~2 performed the standard calibrations---flat-fielding, flux calibration, astrometric alignment with the World Coordinate System (WCS), and photometric zeropoint application---yielding the calibrated (\texttt{\_cal.fits}) products. In Stage~3, the dithered exposures for each filter were combined into final mosaics (\texttt{\_i2d.fits}) using the drizzle algorithm, which resampled the images onto a common grid and improved the effective sampling of the PSF.

The SW images suffer from undersampled PSF, and only the F444W image satisfies the Nyquist sampling criterion. As a result, the PSFs appear asymmetric in all bands except F444W. We therefore measured the flux of AT\,2024tvd as follows: For each NIRCam image (F090W, F150W, F277W, and F444W), we constructed a square cutout centered on the TDE. For the SW images, the cutout size was 28 pixels ($0.87\arcsec$), while for the LW images it was 14 pixels ($0.88\arcsec$).   
The host galaxy light within each cutout was modeled as a smooth two-dimensional low-order polynomial surface using \texttt{astropy} \citep{Astropy_2022}, with a circular region (radius 8 pixels for SW and 4 pixels for LW images) around the TDE masked during the fit. 
We varied the degree of the polynomial to identify the lowest order that minimized residual galaxy flux without overfitting; in practice, a 5th degree polynomial provided the best balance. 
The resulting background model was subtracted from each cutout, leaving the flux from the TDE and any unresolved component at its position, such as an NSC or heated dust. The residual images are displayed in the Appendix.

Fluxes were measured using circular apertures with radii of 5 pixels ($0.16\arcsec$) for the SW images and 2.5 pixels ($0.16\arcsec$) for the LW images. To account for flux falling outside the aperture, we applied position-dependent aperture corrections derived from synthetic PSFs generated with the \texttt{STPSF} package \citep{Perrin_2025_STPSF}. Uncertainties were estimated by combining the statistical (Poisson) noise with a systematic term that captures variations arising from different $1/f$-noise correction methods and from the background modeling process. To quantify the latter, we repeatedly refit the background model after randomly resampling the background pixels and re-measured the source flux each time. The standard deviation of these bootstrap realizations reflects the sensitivity of the flux measurement to the background model. The systematic term dominates the total uncertainty in all four bands. The final photometric measurements are summarized in Table~\ref{tab:nircam_flux} in the Appendix.

\subsection{\textit{JWST} NIRSpec}

The IFU field of view (FoV) is $3\arcsec \times 3\arcsec$, which was centered between the TDE and the galaxy nucleus to ensure both regions were included. Three high-resolution grating-filter combinations---140M/F100LP, G235H/F170LP, and G395H/F290LP---were used to cover the 0.97--5.27~$\mu$m spectral range at a resolution of $R \approx 2700$ (corresponding to a velocity resolution of $\sim$100~km\,s\(^{-1}\)), with small gaps in wavelength coverage at the chip gaps. With a spatial resolution of $0.1\arcsec$ per element, NIRSpec IFU was able to spatially resolve the 0.9\arcsec\ offset between the TDE and nucleus. The \texttt{NRSIRS2} readout pattern and a 4-point dither were used for better sampling of the PSF, and to mitigate any variable detector bias level and 1/$f$ noise. The total exposure times for the 3 grating-filter combinations were 2393, 3560, and 4435~s, respectively.

The NIRSpec IFU data were reduced using the standard \textit{JWST} Science Calibration Pipeline (Version~1.18.0; \citealp{Bushouse_2023_pipeline}) with the Calibration Reference Data System (CRDS) context \texttt{jwst\_1364.pmap}. All processing steps were carried out using this context. The reduction followed the standard three-stage workflow. Stage~1 applied the detector-level corrections---reference pixel, bias, and dark current subtraction, as well as non-linearity and flat-field corrections---before converting the raw data into count-rate images. Stage~2 performed wavelength and flux calibration, background subtraction, and astrometric alignment by assigning WCS information to each exposure, producing 2D spectral images. In Stage~3, the dithered exposures for each grating were combined into fully calibrated 3D spectral cubes (\texttt{*\_s3d.fits}) using the drizzle algorithm to resample the data onto a common spatial and spectral grid.

We executed the full reduction sequence twice: once with the source type set to \texttt{POINT} and once with the source type set to \texttt{EXTENDED}. The \texttt{POINT}-source calibration cube was subsequently used to extract the TDE spectrum (see below), ensuring accurate flux calibration for a compact source, while the \texttt{EXTENDED}-source cube provided optimal calibration and spatial registration for analyzing the host galaxy’s nuclear kinematics. All three final cubes showed low-level sinusoidal patterns in their single-spaxel spectra, caused by resampling noise from the undersampled PSF. These spectral ``wiggles'' were corrected using the \texttt{WICKED} algorithm \citep{Dumont_2025_wicked}.

\begin{figure*}
    \centering
    \includegraphics[width=\textwidth]{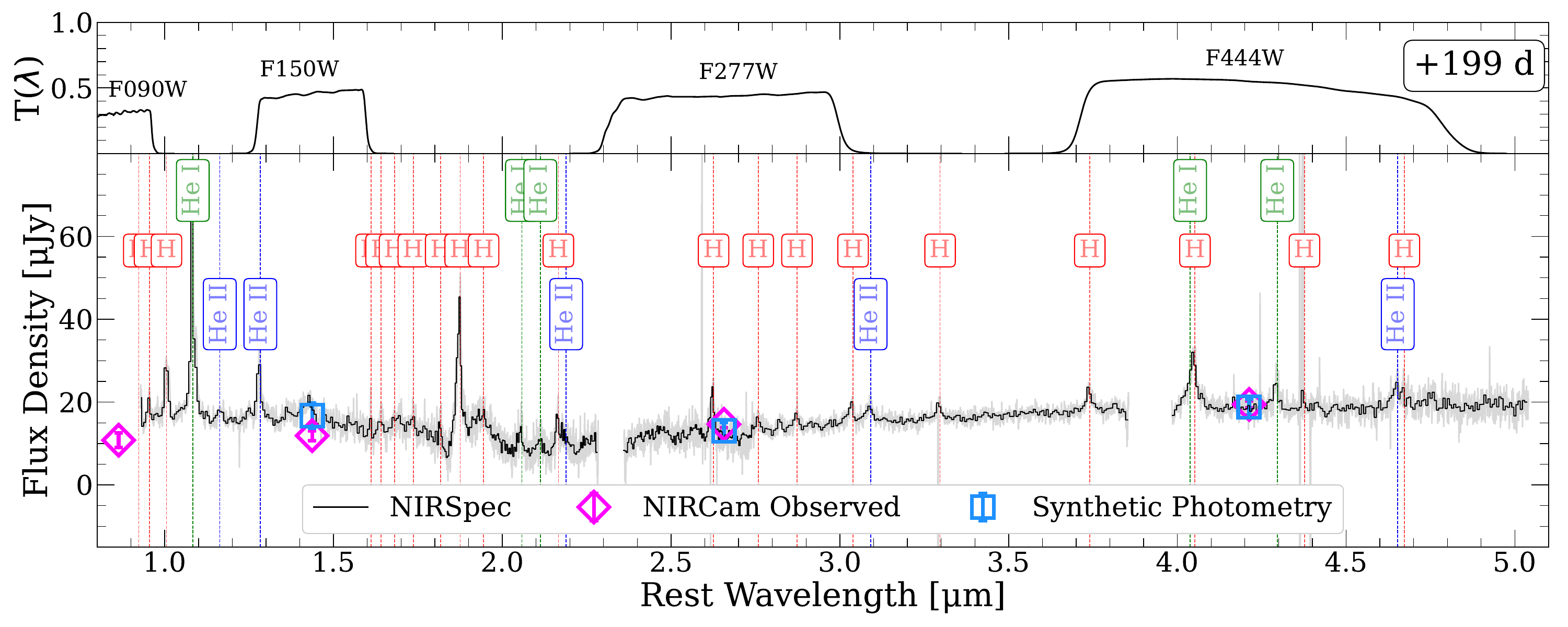}
    \caption{\textit{Upper:} \textit{JWST}/NIRCam filter transmission curves. \textit{Lower:} Infrared spectrum of AT\,2024tvd at phase $+199$ days from \textit{JWST}/NIRSpec, with NIRCam photometry (magenta diamonds) and synthetic photometry derived from the spectrum (blue squares). The observed spectrum is shown in gray, with the black curve displaying the same data rebinned by a factor of 5 for clarity.  Gaps correspond to detector chip gaps in the high-resolution gratings.  Prominent H and He emission lines are indicated.}
    \label{fig:infrared_spectrum}
\end{figure*}
\vspace{1in}
\begin{figure*}
    \centering
    \includegraphics[width=\textwidth]{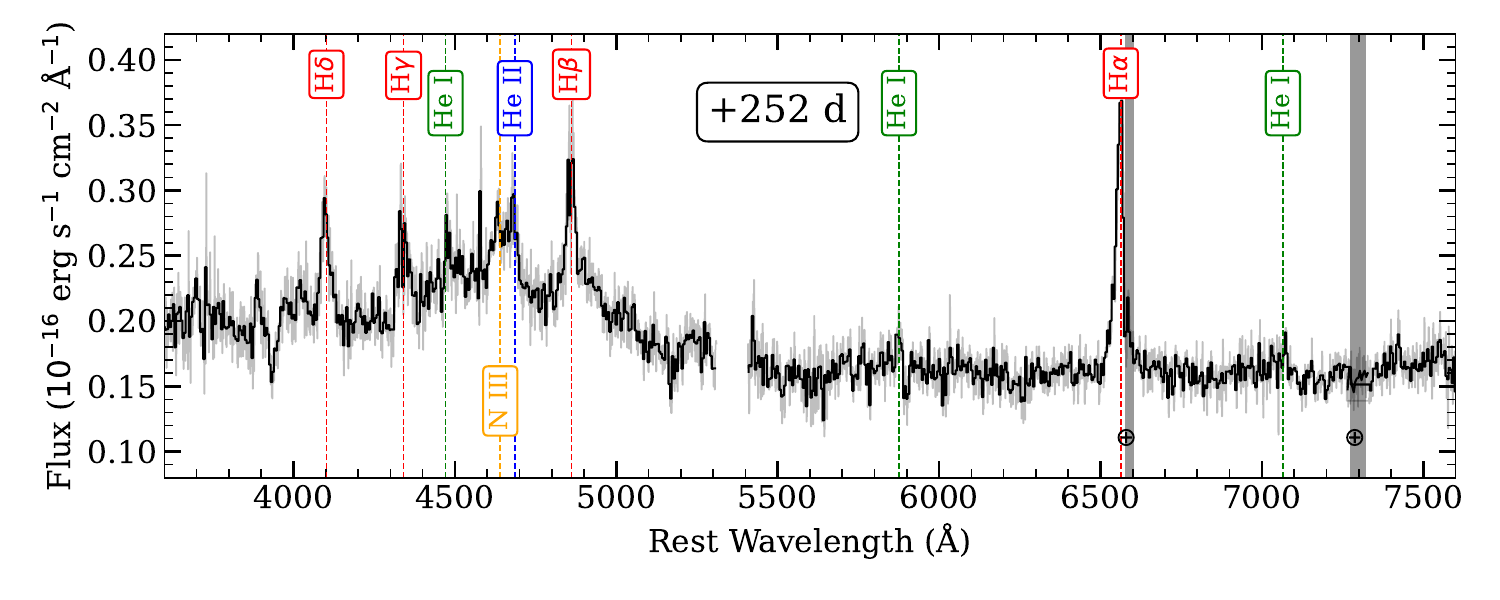}
    \caption{Optical spectrum of AT\,2024tvd at phase $+252$ days observed with KCWI. The observed spectrum is shown in gray, while the overlaid black spectrum has been rebinned by a factor of 5 for clarity. Prominent H and He emission lines are marked. The N\,III Bowen complex is also identified. 
}
\label{fig:optical_spectrum}
\end{figure*}
\vspace{-1in}

The infrared spectrum of AT\,2024tvd, presented in Figure \ref{fig:infrared_spectrum}, was extracted from the NIRSpec IFU datacubes. Following a similar procedure as done for NIRCam photometry, for each wavelength slice, we constructed a square cutout centered on the TDE with a size of 12 pixels ($1.2\arcsec$). The host galaxy light within this cutout was modeled as a smooth two-dimensional polynomial surface using \texttt{astropy}, masking out a circular region of radius 3 pixels ($0.3\arcsec$) centered on the TDE. The polynomial degree was varied to identify the lowest order that minimized the residual galaxy flux without overfitting; in practice, a 5th-degree polynomial provided the best result. This background model was then subtracted from the cutout, leaving only the TDE flux. 
The TDE flux was measured by summing within a circular aperture of radius 2 pixels ($0.2\arcsec$). A wavelength-dependent aperture correction was applied to account for flux outside the aperture, derived from synthetic PSF datacubes generated with the \texttt{STPSF} package. The spectrum was converted from vacuum to air wavelengths following \citet{Morton_1991_vac_air}, and then de-redshifted to the host rest frame at $z=0.04494$.

\subsection{Keck Cosmic Web Imager}

The optical spectrum of AT\,2024tvd is shown in Figure \ref{fig:optical_spectrum}. AT\,2024tvd was observed on 2025 May 26 (UT; MJD 60821), corresponding to phase $+252$~days, with the Keck Cosmic Web Imager (KCWI; \citealp{Morrissey_2018_KCWI}) on the Keck II telescope atop Mauna Kea in Hawai'i. Two pointings were obtained: one capturing the TDE and host galaxy nucleus, and a second that also includes the nearby companion galaxy. Observations were taken in both the blue and red channels, providing usable wavelength coverage from 3500--8000~\AA. The medium slicer was employed in combination with the BL and RL gratings, yielding a spectral resolution of $R \sim 1800$ in the blue and $R \gtrsim 1000$ in the red. The field of view was $16\arcsec \times 20\arcsec$ with a spatial sampling of $0.70\arcsec$ per spaxel. The seeing was $\sim$0.6\arcsec\ at an airmass of $\sim 1.05$, and the observing conditions were excellent. The total exposure time on the blue side was 2000~s, while the red side was observed in 6 individual exposures of 300~s each to mitigate the impact of cosmic-ray hits. The data were reduced and assembled into IFU datacubes using the automated \texttt{KCWI\_DRP} pipeline \citep{Neill_2023_KCWIDRP}. The two pointings were combined into a single IFU datacube for each of the blue and red wavelength arms using a custom implementation of the drizzle algorithm \citep{Fruchter_2002_Drizzle}.

A similar extraction procedure was applied to the KCWI IFU datacubes. The host galaxy was modeled with a two-dimensional polynomial, and the TDE flux was measured within a circular aperture with a radius equal to $2\times$ full-width at half maximum (FWHM) of the seeing disk. A flux correction\footnote{Under seeing-limited conditions, if the PSF is approximated by a \citet{Moffat_1969_psf} function, an aperture with a radius equal to the twice the FWHM of the PSF encloses approximately 95\% of the total flux. Although the FWHM of the seeing disk is wavelength dependent, varying approximately as $\lambda^{-0.2}$ under Kolmogorov turbulence, the resulting change across the 3000--8000~\AA\ range is modest compared to other calibration uncertainties \citep{Fried_1966_Kolmogorov}. We therefore applied a single correction factor for simplicity, noting that any residual wavelength dependence of enclosed flux is at the $\lesssim 5$--10\% level and does not affect our conclusions.} of 1/0.95 was applied. Flux calibration was performed with the \texttt{UCSC Spectral Pipeline}\footnote{\url{https://github.com/msiebert1/UCSC_spectral_pipeline}}, using standard stars Feige\,110 and BD\,+28\,4211 observed with KCWI with the same instrument setup.

\subsection{Public Swift and ZTF data}

To fully characterize the SED of AT\,2024tvd from the X-ray to the infrared, we incorporated publicly available \textit{Swift} and ZTF observations obtained within $\pm5$ days of the \textit{JWST} observations. This ensured a quasi-simultaneous multiwavelength measurement of the SED. The \textit{Swift} XRT data were reduced following standard procedures using the \texttt{HEASoft} software package (v6.35), including the tasks \texttt{xrtpipeline} and \texttt{xrtproducts}. Source spectra were extracted using a circular aperture of radius $40\arcsec$, while background spectra were extracted from nearby source-free regions. Response matrices and ancillary response files were generated with the standard CALDB calibration files, and spectral fitting accounted for Galactic absorption along the line of sight with a column density $N_\mathrm{H} = 4.4 \times 10^{20}$ cm$^{-2}$ \citep{2016_HIPI4}.  

The \textit{Swift} UVOT data in the M2 and W2 filters were reduced with the standard \texttt{uvotsource} task, using a 5\arcsec\ source aperture and a larger nearby background region, with zeropoints from the CALDB database. The host-galaxy SED model---adopted from \citet{Yao_Yuhan_2025_24tvd}---was used to subtract the stellar component from the UVOT data. The UVOT fluxes and the ZTF $r$- and $g$-band fluxes were corrected for host galaxy extinction assuming $E(B-V) = 0.043 $~mag \citep{Schlafly_etal_2011}, adopting the \citet{Cardelli_etal_1989} extinction law with $R_V = 3.1$.

\section{Analysis and Results}
\label{sec:analysis_results}

\subsection{Optical and IR spectra of AT\,2024tvd}

Both the optical and infrared spectra of AT\,2024tvd show prominent H, He\,I, and He\,II emission lines, placing it in the TDE-H+He category of the TDE classification scheme of \citet{vanvelzen21_TDE_tdes}---consistent with the findings of \citet{Yao_Yuhan_2025_24tvd}. In the optical, a broad flux excess between 4500--5000~\AA and N III lines are evident. This corresponds to the Bowen fluorescence complex, which is indicative of fluorescence from strong EUV/soft X-ray emission reprocessed in dense---possibly, CNO-cycle-enriched---gas near the SMBH \citep{Onori_2019_bowen}. 
The infrared spectrum shows a suite of hydrogen recombination lines from the Paschen, Brackett, Pfund and Humphreys series, along with a particularly strong He\,I~$\lambda1.083~\mu$m line, and several He\,II lines, further confirming the ionized gas conditions. The infrared continuum initially declines with wavelength, reaching a minimum near $2.3\,\mu$m, before rising again at longer wavelengths. No evidence of dust precursors, such as broad CO overtone features, is detected in the infrared, suggesting that no new dust formation is occurring at the time of observation (e.g., \citealp{Derkacy_2025_23ixf}).

Figure \ref{fig:infrared_spectrum} also compares the observed NIRCam photometry with synthetic photometry derived from the infrared spectrum, for all filters except F090W whose transmission function does not significantly overlap with the spectral coverage. The synthetic photometry was computed using  
\begin{equation}
    F_{\nu,\,\mathrm{syn}} = \frac{\int F_{\nu}(\lambda)\, T(\lambda)\, \frac{d\lambda}{\lambda}}{\int T(\lambda)\, \frac{d\lambda}{\lambda}},
\end{equation}  
where $F_{\nu}(\lambda)$ is the flux density of the spectrum and $T(\lambda)$ is the transmission curve of the filter. The uncertainty on the synthetic photometry was estimated as the standard deviation (excluding the emission lines) of the spectrum within the wavelength range of each filter, scaled by the effective filter width, thereby capturing both statistical noise and residual systematics. This comparison was performed to check for consistency between the imaging and spectroscopic data, and we find good agreement across all overlapping filters.

\subsection{Kinematics: line-of-sight velocity, dispersion, metallicity and stellar age maps}

\begin{figure*}
    \centering
    \includegraphics[width=\textwidth]{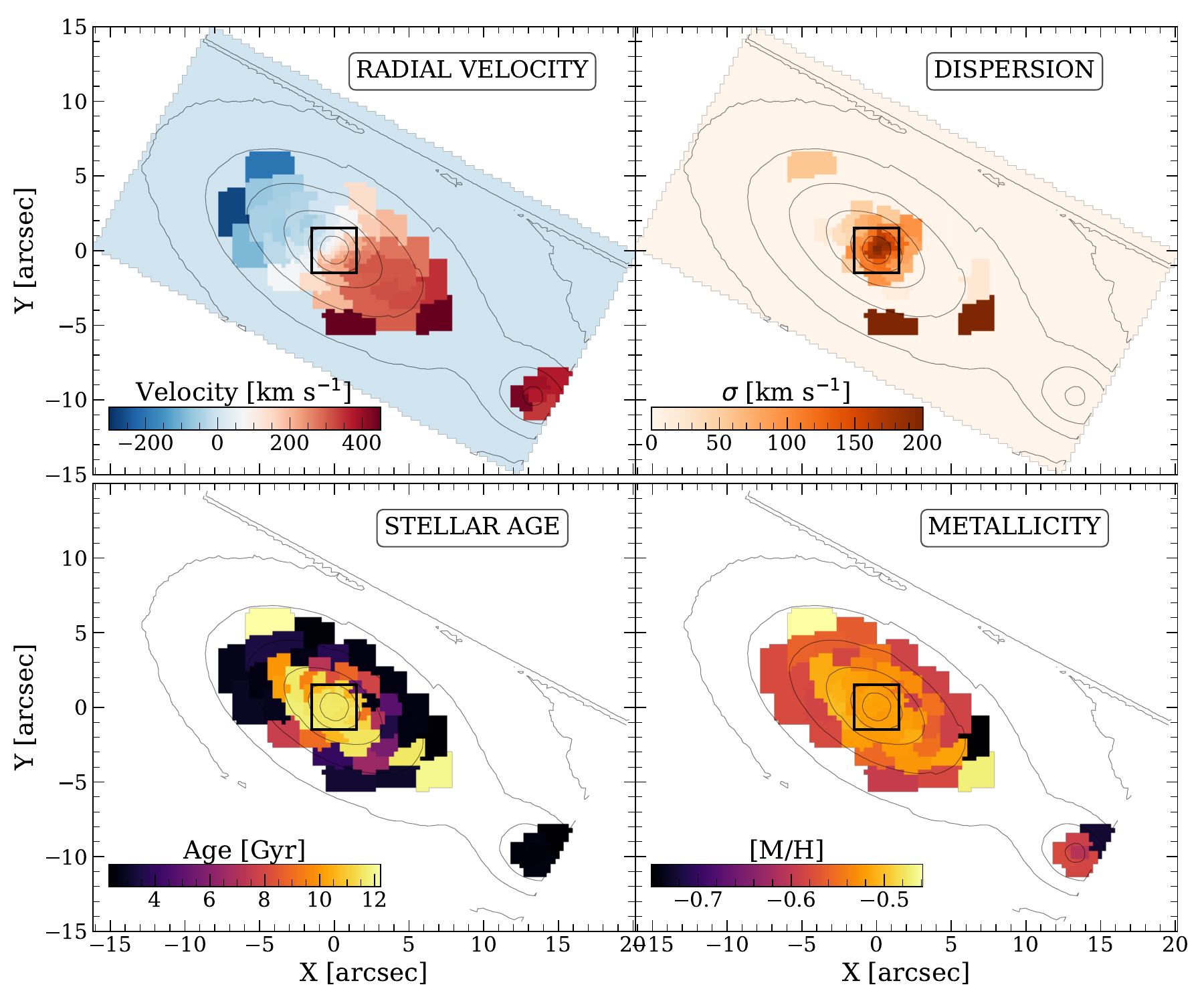}
    \caption{KCWI kinematic maps of the host galaxy of AT\,2024tvd and its nearby companion. Flux contours are overlaid, and the inner black box indicates the \textit{JWST}/NIRSpec field of view.}

    \label{fig:kcwi_kinematics}
\end{figure*}

\begin{figure*}
    \centering
    \includegraphics[width=\textwidth]{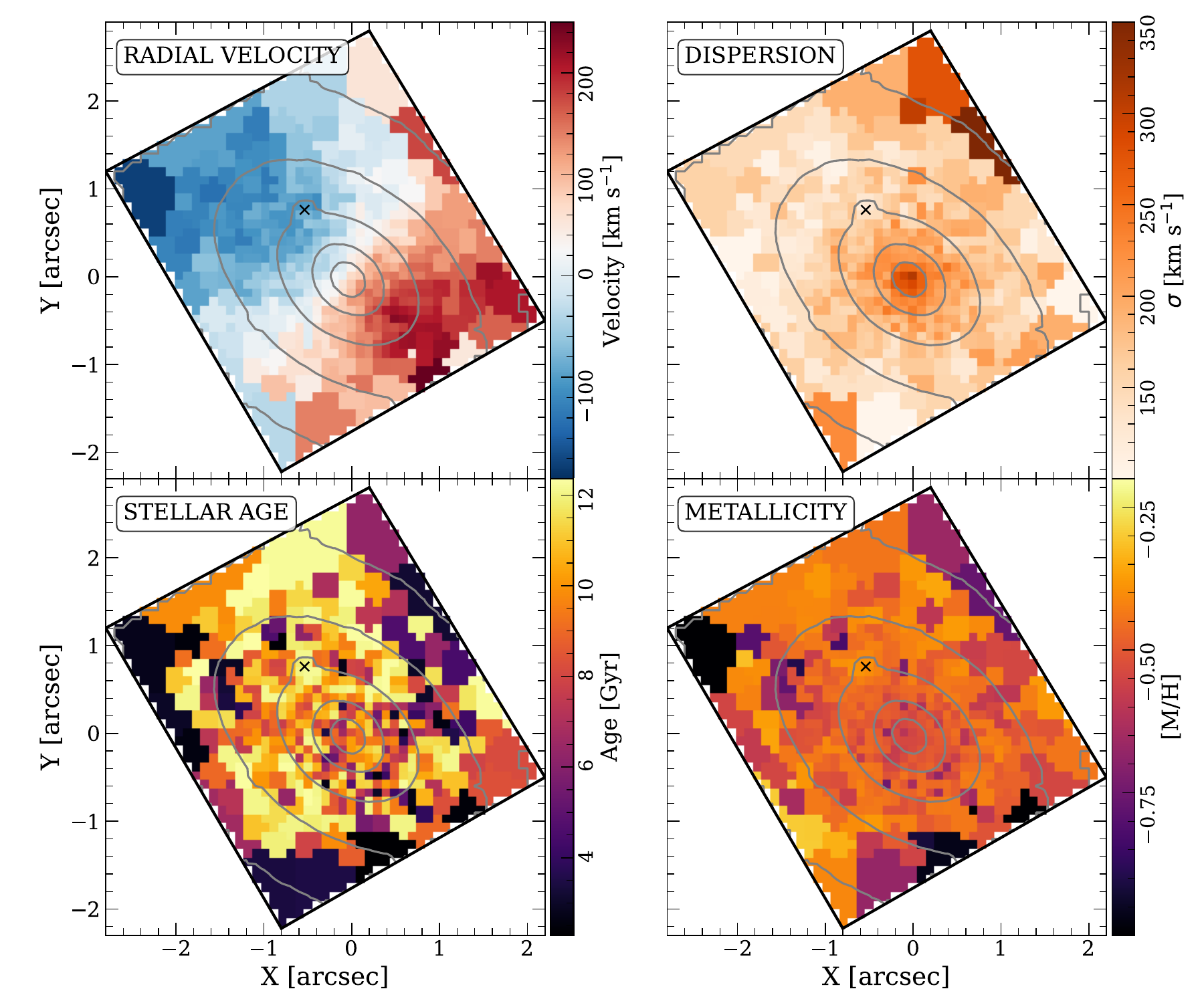}
    \caption{\textit{JWST}/NIRSpec kinematic maps of the host galaxy of AT\,2024tvd. Flux contours are overlaid, and the location of the TDE is marked with an ``X''.
}
    \label{fig:jwst_kinematics}
\end{figure*}

The stellar kinematics of the host galaxy were measured from the KCWI IFU and NIRSpec datacubes---presented in Figures \ref{fig:kcwi_kinematics} and \ref{fig:jwst_kinematics}, respectively---using the penalized pixel-fitting method (\texttt{pPXF}; \citealt{Cappellari_2012_ppxfsoft, Cappellari_2017_ppxfpaper}). Prior to fitting, the cubes were spatially binned using Voronoi tessellation \citep{Cappellari_2003_voronoi} to achieve a minimum signal-to-noise ratio ($S/N$) of $\sim 50$ per bin. The spectra in each bin were logarithmically rebinned in wavelength, which makes velocity shifts correspond to linear shifts in $\rm log~ \lambda$ space. 
We adopted the E-MILES stellar population templates \citep{Vazdekis_2016_emiles} for the fits. E-MILES provides empirical, high-spectral-resolution templates with extended wavelength coverage into the near-infrared up to $5~\mu$m, which is necessary for fitting \textit{JWST} spectra. The template spectra were convolved to match the instrumental resolution of NIRSpec and KCWI prior to fitting.

The \texttt{pPXF} algorithm fits each binned spectrum by linearly combining the template spectra, convolved with a parameterized line-of-sight velocity distribution (LOSVD). In this work, we fit for the first two velocity moments: the recession velocity and the velocity dispersion. Higher Gauss–Hermite moments were not included. To avoid overfitting and to ensure physically smooth variations in the star formation history, we adopted a regularization parameter of \texttt{regul} $=10$. Multiplicative Legendre polynomials were used to account for the continuum shape, while strong emission lines were masked. For each Voronoi bin, the best-fitting LOSVD yields the stellar velocity and velocity dispersion, while the light-weighted combination of templates provides the mean stellar age and metallicity of the population. The product of this procedure is two-dimensional maps of the stellar velocity, velocity dispersion, age, and metallicity across the field.

The KCWI velocity maps cover the full host galaxy and the nearby companion, whereas the \textit{JWST}/NIRSpec IFU data include only the innermost region of the host. The host galaxy is classified as an S0 Lenticular galaxy with both disk and bulge components. The bulge has an effective radius of $R_{\mathrm{e}} = 3.3~\mathrm{kpc}$ ($3.6''$; \citealt{Simard_2011_galcat}), such that the $3''\times3''$ NIRSpec field of view encompasses nearly the entire bulge, while the larger KCWI field of view captures both the bulge and the extended disk component. 

Clear rotation is visible in the velocity maps from both instruments. While rotation in the galactic disk is expected, the presence of strong rotational support in the bulge suggests that the bulge has not yet fully virialized. The KCWI data yield a bulge velocity dispersion of $\sim200~\mathrm{km\,s^{-1}}$, consistent with the value measured from the SDSS spectrum of the host \citep{Gunn_2006_SDSS, Ahumada_2020_SDSS}. The stellar population in the central regions is old ($\sim$12--13 Gyr) with a metallicity of $\log(Z/Z_\odot) \approx -0.5$, both in agreement with the host SED-based measurements reported by \citet{Yao_Yuhan_2025_24tvd}. At the higher spatial resolution of the \textit{JWST} IFU, the TDE is clearly resolved from the nucleus. No significant differences in velocity, velocity dispersion, stellar age, or metallicity are detected at the location of the TDE compared to the surrounding bulge, and there is no evidence for stellar disturbances in the kinematic maps around the TDE.

The sphere of influence of a black hole is defined as
\begin{equation}
    r_{\mathrm{soi}} = \frac{G M_\bullet}{\sigma^2},
\end{equation}
where $M_\bullet$ is the black hole mass and $\sigma$ is the stellar velocity dispersion. Using the $M_\bullet$--$\sigma$ relation \citep[e.g.,][]{McConnell_2013_Msigma, Kormendy_2013_Msigma} and the measured velocity dispersions of $\sigma \approx 150$--$200~\mathrm{km\,s^{-1}}$ (at the TDE location and in the bulge), the expected black hole masses are $M_\bullet \approx (1$--$2)\times10^8~M_\odot$, corresponding to spheres of influence of $\sim15$--$20$~pc ($\theta_{\mathrm{soi}} \approx 0.02''$ at $z=0.045$). Given the $\sim0.7''$ PSF and $0.1''$ pixel scale of the \textit{JWST} IFU data, these spatial scales are far below the instrumental resolution, implying that neither the off-nuclear nor the central black hole can be dynamically resolved. Consequently, the IFU kinematics can only place a loose upper limit of $M_\bullet \lesssim 10^9~M_\odot$. Such an upper limit is not astrophysically useful for TDEs, since all plausible TDE-producing black holes have $M_\bullet \lesssim 10^8~M_\odot$ \citep[e.g.,][]{Hills_1975, Rees_1988, 2012_Macleod, Gezari_2021_tderev}. The mass of the TDE-causing black hole must therefore be inferred through other means, such as SED modeling, light-curve fitting, or scaling relations with other observables. For the central black hole, the best available mass estimate, $M_\bullet \approx 2 \times 10^8~M_\odot$, remains that derived from the $M_\bullet$--$\sigma$ relation of the host bulge \citep{Yao_Yuhan_2025_24tvd}.

The companion galaxy lies on the redshifted side of the host’s rotation field, and its systemic velocity is also redshifted. This is consistent with the motion of stars in the adjacent region of the host. While this alignment does not by itself imply a dynamical connection, it is suggestive of a prograde orbital configuration if the two galaxies have interacted \citep[e.g.,][]{Barnes_1992_galmergrev, Naab_2003_galmerg}. Additionally, the companion galaxy appears to be a younger spiral system with a stellar age of $<4$ Gyr and a lower metallicity than the host of AT\,2024tvd.

\subsection{Spectral Energy Distribution Modeling of AT~2024tvd}

\begin{figure*}
    \centering
    \includegraphics[width=\textwidth]{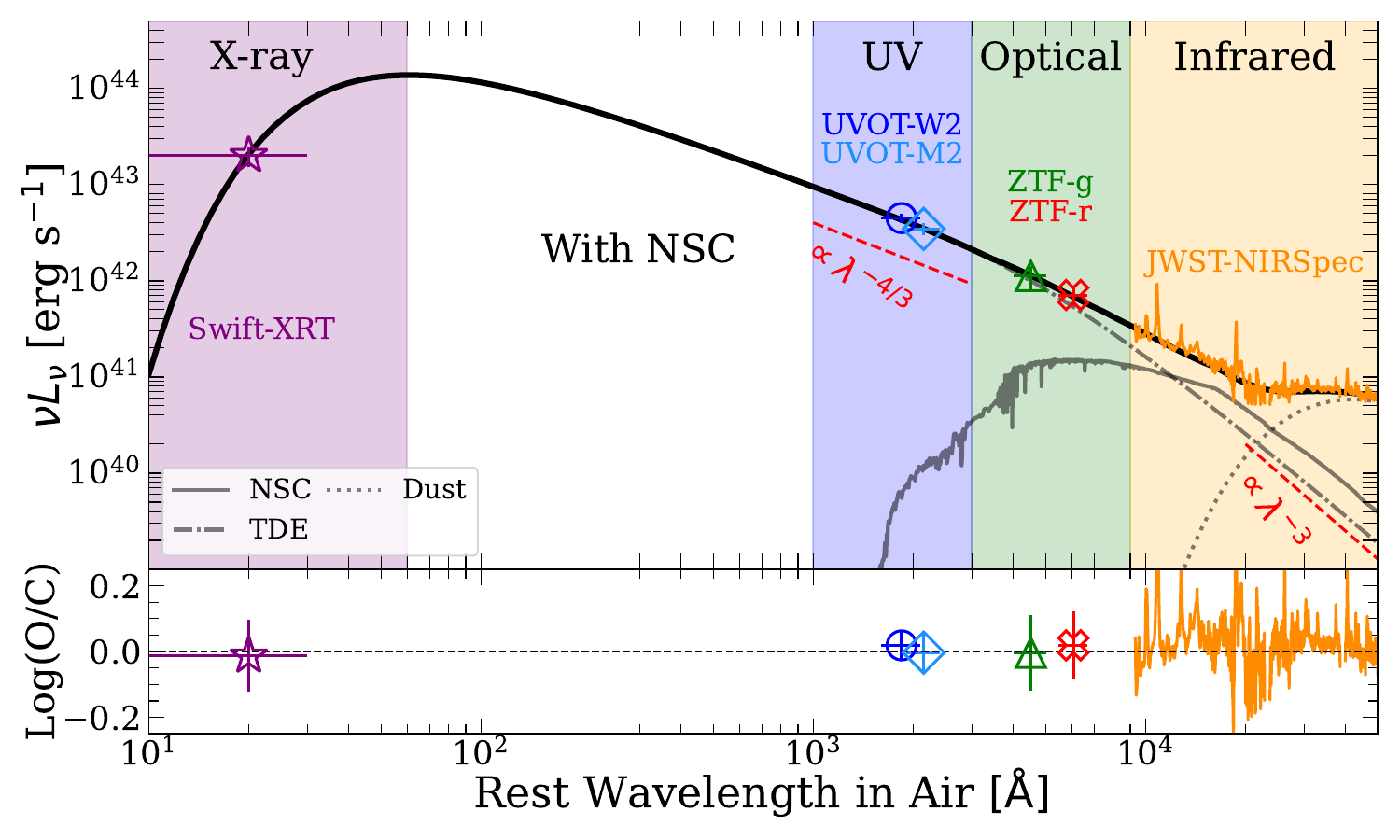}
    
    \includegraphics[width=\textwidth]{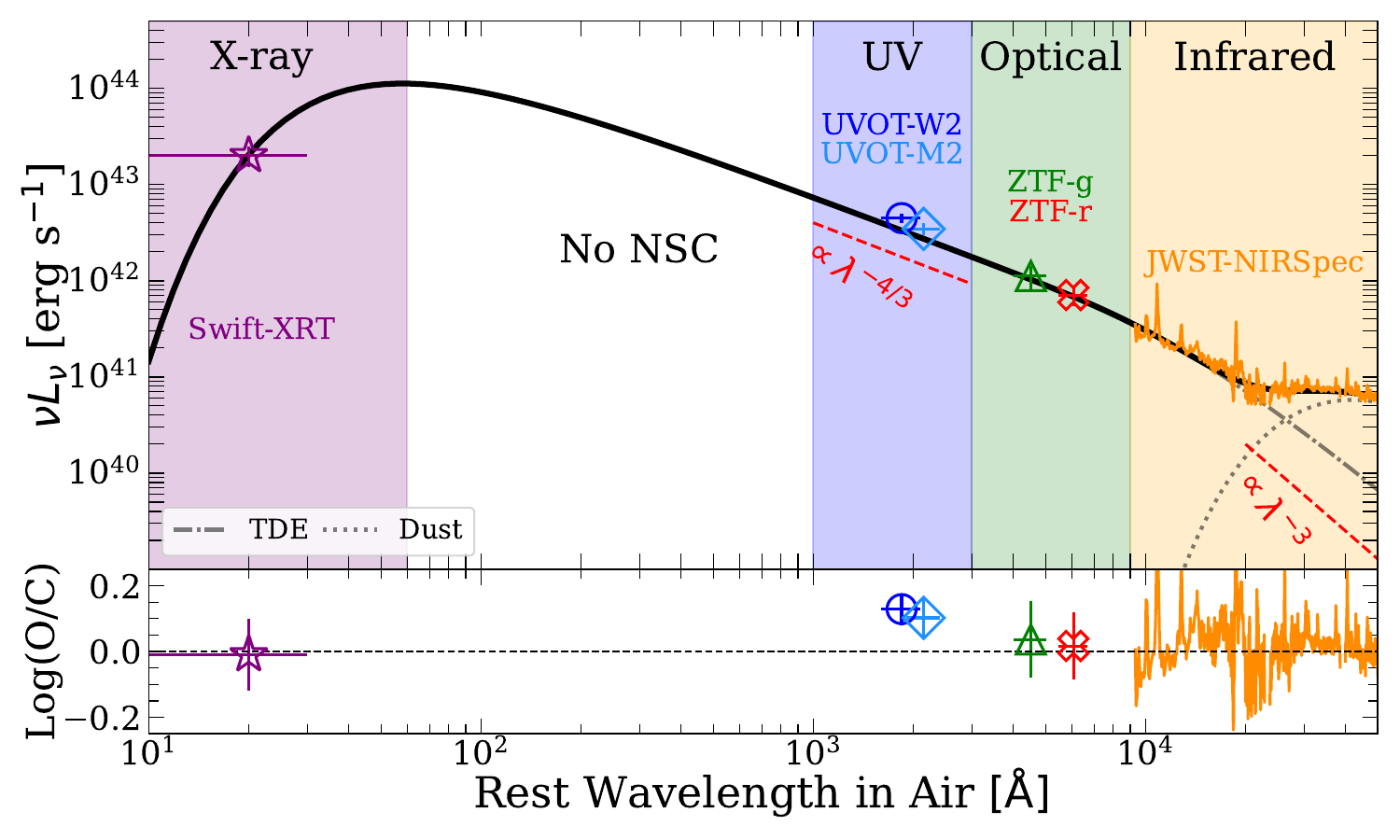}
    \caption{\textit{Upper:} SED modeling of AT\,2024tvd including the NSC component. The gray curves show the individual model components (accretion disk, nuclear star cluster, and dust emission), while the solid black curve shows the best-fit combined SED. The residuals (logarithm of observed minus calculated) are also shown. In the UV–optical regime, the multicolor blackbody disk follows the characteristic $\lambda^{-4/3}$ slope, whereas at longer wavelengths the finite disk size leads to a steeper infrared decline of $\lambda^{-3}$. \textit{Lower:} Same as the upper panel but without the NSC component. Note that the residuals are worse, particularly in the UV bands. 
}
    \label{fig:SED_NSC}
\end{figure*}

We performed a comprehensive SED analysis (Figure \ref{fig:SED_NSC}) of AT\,2024tvd using a combination of stellar population (NSC), accretion disk (multicolor blackbody), and dust emission models. The goal was to disentangle the contribution of these components to the overall SED and to constrain the physical parameters of the accretion flow, the NSC (if present) and dust.

The observed data from \textit{Swift}/XRT (X-ray), \textit{Swift}/UVOT (UV), ZTF-$g$ and ZTF-$r$ (optical), and \textit{JWST}/NIRSpec (infrared) were first shifted to the rest frame using the measured redshift of the host galaxy. For the \textit{JWST}/NIRSpec spectrum, the wavelength scale was additionally converted from vacuum to air values to ensure consistency across all datasets. All fluxes were converted into luminosity ($\nu L_{\nu}$ [erg s$^{-1}$]), and for the SED fitting we worked in log space to ensure numerical stability given the large dynamic range of the data ($\sim 10^{40}$--$10^{43}$~erg~s$^{-1}$).

Only the continuum of the \textit{JWST} spectrum was used in the fitting: spectral lines and their wings up to $2\times$FWHM were masked, and the continuum was binned above the spectral resolution to minimize correlated noise introduced by the line-spread function and astrophysical line broadening. This ensures that the sampled data points are effectively independent, allowing for a simpler and more accurate treatment of the uncertainty. Since our modeling does not include emission lines, we note that the \textit{Swift}/UVOT bands used here do not overlap with any prominent lines observed in the UV spectrum obtained with \textit{HST} \citep{Yao_Yuhan_2025_24tvd}. On the other hand, the ZTF-$g$ and ZTF-$r$ bands include flux from some emission features. However, based on convolving the ZTF filter transmission curves with the KCWI optical spectrum, we estimate that neglecting this excess introduces only minor biases of $\sim$9\% and 5\%, respectively---smaller than the statistical uncertainties of 26\% and 22\%. We therefore do not adjust the ZTF fluxes, instead treating the emission-line contribution as a systematic uncertainty added in quadrature with the statistical error.

%\clearpage

The stellar component was modeled using pre-computed Flexible Stellar Population Synthesis (\texttt{FSPS}) %single stellar population 
grids \citep{Conroy_etal_2009_FSPS, Conroy_Gunn_2010_FSPS}, spanning ages of $1$--$14$~Gyr and metallicities in the range $-2.0 < \log (Z/Z_\odot) < +0.5$. A pure single-stellar population model was used with a \citet{2001_Kroupa} initial mass function. To accelerate the inference, we interpolated within this grid rather than calling \texttt{FSPS} at each likelihood evaluation.

For modeling the accretion disk emission, we adopted a standard multi–color blackbody prescription, in which the disk is geometrically thin, optically thick, and radiates locally as a blackbody \citep{Shakura_1973_accretion, Novikov_1973_accretion}. The effective temperature profile follows the Shakura–Sunyaev/Novikov–Thorne solution with a zero–torque inner boundary at the innermost stable circular orbit (ISCO). This ensures that the flux smoothly vanishes at the inner edge rather than diverging \citep{Page_1974_fr, Zimmerman_2005_ezdiskbb}. The emission was integrated over concentric annuli assuming axisymmetry. We neglected relativistic transfer effects (e.g., gravitational redshift, Doppler boosting, and light bending), which is reasonable given 
our focus on the optical–IR regime where such effects are small \citep[e.g.,][]{Cunningham_1975_accretion, Li_2005_MCD}. 
We also imposed an explicit outer radius that sets the low–frequency cutoff of the spectrum. This model is parametrized by the black hole mass $\log (M_{\rm \bullet}/M_\odot)$, mass-accretion rate $\log (\dot{M}/M_{\odot}~\rm yr^{-1})$, and an outer-disk radius $\log(r_{\rm out}/r_g)$. Here, $r_g$ is the gravitational radius given by $r_g \equiv GM_{\bullet}/{c^2}$. 

Finally, the dust emission was modeled with a simple blackbody, characterized by a peak normalization $\log (A_{\rm peak}/\rm erg~s^{-1})$ and dust temperature $T_{\rm dust}$ [K]. We also included an additional nuisance parameter representing an intrinsic scatter term.

We used the \texttt{emcee} affine-invariant sampler \citep{Foreman-Mackey_etal_2013_emcee} to explore the posterior distribution of the full parameter set. The likelihood function assumed Gaussian errors in log-flux space with the additional intrinsic scatter term. We ran chains with $N_{\rm walkers} = 48$ and a total of $N_{\rm steps} = 10^{5}$ steps, discarding the initial 30\% of samples as burn-in.
The parameters fitted with \texttt{emcee}, together with their adopted priors and summary statistics of the posterior distributions, are provided in Table~\ref{tbl:emcee_results} in the Appendix section.

We explicitly fit two models: the first model included contributions from the TDE accretion disk, dust emission, and an NSC. The second model omitted the NSC, consisting of only the accretion disk and dust components. In the first case, the NSC contribution is well constrained (Figure \ref{fig:SED_NSC}), with the posterior favoring a distinct stellar component in addition to the accretion disk and dust. To assess which of these scenarios is statistically preferred, we compared their Bayesian Information Criterion (BIC; \citealp{Schwarz_1978_BIC}) value, which penalizes the model with higher number of free parameters. The model including the NSC yields $\mathrm{BIC}=-307$, while the model without the NSC yields $\mathrm{BIC}=-298$. Since a lower BIC indicates a better fit, and the difference $\Delta \mathrm{BIC}=9$ constitutes strong evidence in favor of one model over another \citep{kass1995bayes}, we conclude that the data statistically prefer the existence of a nuclear star cluster. We note, however, that at face value, the model without an NSC also provides a reasonable fit to the data.

For the accretion component, our fits constrain the black hole mass, the instantaneous accretion rate, and the outer extent of the accretion disk. 
For the NSC, we obtain an estimate of the stellar mass, although the age and metallicity of the population are only loosely constrained, since our SED modeling relies only on the continuum shape of the \textit{JWST} infrared spectrum. In contrast, the dust component is well characterized: both the dust temperature and its peak luminosity are tightly constrained by the data. We discuss each of these components in detail in Section \ref{sec:discussion}.

\subsection{Unsharp Mask}
\label{sec:unsharp}

To search for faint morphological signatures of galaxy mergers or MBH interaction with the stellar field, we applied an unsharp masking technique to the \textit{JWST}/NIRCam F444W image. Among the four NIRCam images, F444W provides the deepest sensitivity ($\sim 27.5$ mag AB) in the 1000\,sec exposures. Unsharp masking is an image-processing method in which a smoothed version of the original image is subtracted from the data, resulting in enhanced sharp edges and small-scale structures in the image (e.g., \citealt{Auchere_2023_unsharp}). This procedure suppresses the smooth stellar light of the host galaxy while highlighting compact or high-contrast features such as clumps, shells, or tidal tails. 

\begin{figure*}[!htb]
    \centering
    \includegraphics[width=\textwidth]{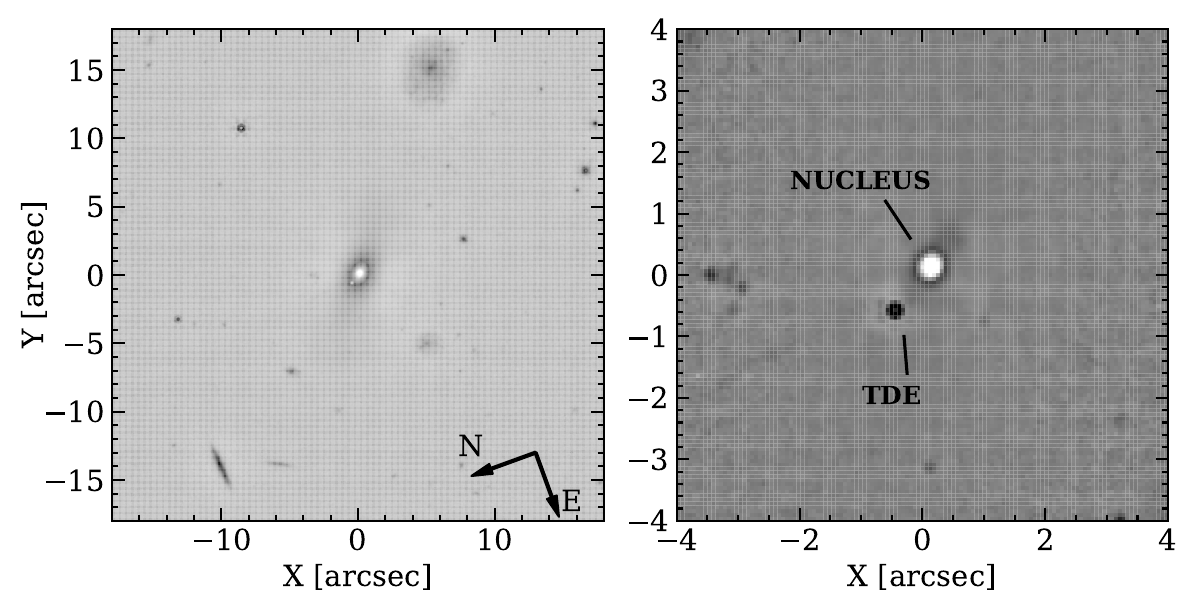}
    \caption{Unsharp-masked \textit{JWST}/NIRCam F444W image. \textit{Left}: A $36\arcsec \times 36\arcsec$ view centered on the host galaxy nucleus.
\textit{Right}: A zoomed-in $8\arcsec \times 8\arcsec$ region showing the off-nuclear TDE just below and to the left of the nucleus. Subtracting the smooth stellar background reveals several faint satellite galaxies in the surrounding field but no evident tidal tails, shells, or other morphological disturbances. }
    \label{fig:jwst_unsharp}
\end{figure*}

The unsharp-masked image of AT\,2024tvd’s field (Figure~\ref{fig:jwst_unsharp}) reveals that the host galaxy is remarkably smooth and featureless, with no evidence for tidal tails, shells, or large-scale stellar disturbances. We therefore find no morphological indication of a recent \textit{major} merger or bulk displacement of stars associated with the offset MBH that produced the TDE. In contrast, the companion galaxy exhibits distinct spiral arms and clumpy star-forming knots. Diffraction spikes from a bright Galactic star are also enhanced, and several faint satellite galaxies, otherwise hidden in the bright host-galaxy light, become visible, pointing to a relatively rich environment. On the other hand, this picture is consistent with a \textit{minor} merger scenario, where large-scale morphological disturbances may not be observed \citep[e.g.,][]{2021_Dodd}.

\section{Discussion}
\label{sec:discussion}

\subsection{SED model parameters}

\textbf{Black hole mass:} Based on our MCMC fits, we infer a black hole mass of $\log (M_{\bullet}/M_{\odot}) = 5.50\pm 0.04$. This value is consistent with the estimate of \citet{Yao_Yuhan_2025_24tvd}, who derived $\log (M_{\bullet}/M_{\odot}) = 6 \pm 1$ from the peak bolometric luminosity. Our result is also not far from the value determined by \citet{Yao_Yuhan_2025_24tvd} using the Modular
Open-Source Fitter for Transients, \texttt{MOSFiT}, TDE model \citep{Guillochon_2018_mosfit, 2019_Mockler}, $\log (M_{\bullet}/M_{\odot}) = 5.89^{+0.15}_{-0.06}$, but differs significantly from the higher mass of $\log (M_{\bullet}/M_{\odot}) = 6.9 \pm 0.5$ inferred from the luminosity of the late-time UV plateau \citep{Mummery_2024_BH-TDE_scaling}. The general consensus, however, is that the offset black hole responsible for the TDE has a mass significantly lower than that expected for the nuclear black hole from independent estimates based on scaling relations such as $M_{\bullet}$--$\sigma$ and $M_{\bullet}$--$M_{\mathrm{gal}}$, which are generally around $M_{\bullet} \approx 10^{8.5}~M_{\odot} $.

For a low-mass black hole (\(M_\bullet \approx 10^{5.5}\,M_\odot\)), the tidal disruption radius for a solar-type star lies well outside the gravitational radius, typically \(R_{\rm t} \sim 100\,R_{\rm g}\) \citep{Rees_1988}. Although general-relativistic precession is weaker at such large distances---potentially making debris circularization less efficient---the observed light curve of AT\,2024tvd rises on a timescale that is consistent with the expectation for efficient circularization \citep{Guillochon2015}: the characteristic fallback time of the most-bound debris, which sets the rise time if the debris promptly forms an accretion disk, is
\begin{equation}
t_{\rm fb} \approx 41~{\rm d}\, M_6^{1/2} r_*^{3/2} m_*^{-1} \beta^{-3}
\end{equation}
\citep{Rees_1988, stone13_frozen_in, Guillochon_Ramirez_2013}.
For \(M_\bullet = 10^{5.5}\,M_\odot\), a solar-type star, and \(\beta \approx 1\), this gives \(t_{\rm fb} \approx 23~{\rm d}\)---consistent with the observed $\sim25$-day rise of AT\,2024tvd’s optical light curve \citep{Yao_Yuhan_2025_24tvd}. Circularization of the stellar debris may be facilitated by stream self-intersections and hydrodynamic shocks (e.g., \citealp{Shiokawa_etal_2015, Lu_Bonnerot_2020}), and could even be made more efficient if residual gas from a previous TDE is present, as might be expected if the TDE rate around this offset MBH is elevated. 

On the other hand, an MBH as massive as \(M_\bullet \sim 10^{8.5}\,M_\odot\) is inconsistent with the observed properties of AT\,2024tvd.  
At such high masses, the tidal radius for a solar-type star lies near or within the MBH's event horizon (e.g., \citealp{Kesden2012}), implying that the star would be swallowed whole rather than producing an observable flare \citep{Rees_1988,2012_Macleod,Law-Smith2017}. Even for a rapidly spinning black hole, the disruption would occur only marginally outside the innermost stable circular orbit, making a luminous TDE highly unlikely. In addition, the expected fallback time of the most-bound debris scales as \(t_{\rm fb} \propto M_\bullet^{1/2}\).  
For \(M_\bullet = 10^{8.5}\,M_\odot\), this gives \(t_{\rm fb} \approx 700\)~days for a solar-type star and a grazing encounter with \(\beta \approx 1\) \citep{stone13_frozen_in, Guillochon_Ramirez_2013}. To reproduce the observed \(\sim25\)-day rise time, the disruption would need to be extremely deep (\(\beta \gtrsim 3\)), corresponding to a pericenter well inside the event horizon, which is an unphysical scenario. One might consider whether a more extended star, such as a red giant, could reconcile the short timescale. However, since \(t_{\rm fb} \propto R_*^{3/2}\), a larger stellar radius would only \textit{increase} the fallback time, yielding a slower rise than observed.  
Together, these arguments make it highly unlikely that a very massive (\(\gtrsim10^8\,M_\odot\)) black hole caused the TDE.

\textbf{Accretion rate:} The mass accretion rate from our modeling is $\log( \dot{M}/M_\odot\,{\rm yr}^{-1}) = -1.22\pm0.04$. This represents an instantaneous measurement of the accretion rate, since the SED corresponds to a quasi-simultaneous snapshot at $+200$ days after the optical peak. If the disrupted star had $M_\star \simeq 1\,M_\odot$, this would imply an order-of-magnitude accretion lifetime of $M_\star/\dot{M} \approx 17$ yr. However, accretion will likely persist much longer at a lower rate, as shown in theoretical work and long-term observational monitoring of TDEs \citep[e.g.,][]{Mummery_2025_TDEdisktheory, vanVelzen_2019_UVTDEdisk, Guolo_2025_disksize, 2022_Thomsen, Dai_etal_2018, 2017_Auchettl}. 

The inferred accretion rate is super-Eddington: for a black hole mass of $3\times10^{5}\,M_\odot$ and an assumed radiative efficiency of $\eta = 0.1$, the rate corresponds to $\sim 10$ times the Eddington limit \citep{2009_Ramirez-Ruiz, 2012_DeColle}. Under such conditions, the accretion flow is expected to be a geometrically thick, radiation-pressure-dominated disk, accompanied by powerful outflows \citep{Strubbe_etal_2009, Sadowski_2014_disksim, 2022_Thomsen, Dai_etal_2018}. An observational consequence of such outflows may be connected to the second radio flare observed in AT\,2024tvd at $\sim 200$ days \citep{Sfaradi_2025_24tvdradio}. It is therefore plausible that the super-Eddington accretion inferred at this epoch reflects, at least in part, the outflowing material responsible for the observed radio flares in AT\,2024tvd.

\textbf{Disk's outer radius:} 
The outer radius of a multi–color blackbody (MCB) disk sets the long–wavelength edge of the SED: in the intermediate UV–optical regime the MCB slope follows the thin–disk asymptote, $\nu L_{\nu}\propto \nu^{4/3}\propto \lambda^{-4/3}$, while at wavelengths longer than those emitted by the coolest annulus ($\lambda \gtrsim hc/kT[R_{\rm out}]$) the spectrum rolls over and steepens ($\nu L_{\nu}\propto \lambda^{-3}$) toward the Rayleigh–Jeans limit, producing an IR cutoff \citep{Shakura_1973_accretion, Novikov_1973_accretion}. For AT\,2024tvd we infer $R_{\rm out}\simeq 10^{3.82}\,r_g\approx 6\times10^{14}\,{\rm cm}$. This value is comparable to the blackbody radii commonly derived for optical TDE emission ($\sim 10^{14}$–$10^{15}\,{\rm cm}$; \citealt{Yao_Yuhan_2023_TDEs}), suggesting that the outer accretion flow and any reprocessing layer may be closely coupled. This radius is $\sim 10^2$ times larger than the tidal radius (for a solar–type star), implying substantial viscous spreading of the disk. Alternatively, this is also consistent with an extended super-Eddington photosphere dominated by outflows \citep{Dai_etal_2018, 2022_Thomsen, Roth_etal_2016}. 

A simple viscous estimate $t_{\rm visc}\!\approx\!\alpha^{-1}(R/H)^2\Omega_K^{-1}$ gives $t_{\rm visc}(R_{\rm out})\!\sim\!{\rm few}$ years for a moderately thick flow ($H/R\!\approx\!0.3$, $\alpha\!\approx\!0.1$). But it can be as short as $\sim 200$ days if the flow is very thick ($H/R\!\approx\!1$) and efficiently transporting angular momentum. Thus, achieving $R_{\rm out}\!\approx\!6\times10^{14}$~cm by $+200$~d is plausible in a super–Eddington, wind–supported disk, but would be on the large side for a thin, cool disk \citep[e.g.,][]{Strubbe_etal_2009, Dai_etal_2018, Lu_Bonnerot_2020}. This most likely indicates that the accretion disk at $+200$\,d is still transitioning from a reprocessing dominated flow to an unobscured relativistic thin disk \cite{Mummery_2024_BH-TDE_scaling}.

\textbf{Mass of the nuclear star cluster:} The inferred NSC mass from our modeling is $\log (M_{\rm NSC}/M_{\odot}) = 7.97^{+0.16}_{-0.26}$. In the scenario where the host nucleus already contains a massive black hole of $M_{\bullet,\mathrm{host}}\!\approx\!10^{8.5}\,M_\odot$, and the  off–center TDE is powered by a secondary black hole of $M_{\bullet,\mathrm{sat}}\!\approx\!10^{5.5}\,M_\odot$ brought in through a galaxy merger, the inferred NSC mass of $M_{\star,\mathrm{NSC}}\!\approx\!10^{8}\,M_\odot$ at the TDE location is high, but consistent with a nucleated satellite. Observationally, NSCs span $10^{5}$–$10^{8}\,M_\odot$ and broadly correlate with host stellar mass \citep{Boker_2004_NSC, Neumayer_2020_NSCrev}. An NSC of $10^{8}\,M_\odot$ typically corresponds to a progenitor galaxy of $\sim 10^{10}\,M_\odot$ (for canonical NSC mass fractions of $0.1$–$1\%$; \citealp{Neumayer_2020_NSCrev}).

Relative to the inferred offset black hole mass of $M_{\bullet}\!\approx\!10^{5.5}\,M_\odot$, the ratio $M_\bullet/M_{\star,\mathrm{NSC}}\!\simeq\!10^{-2.5}$ is low but consistent with empirical trends in which NSCs often dominate the nuclear mass budget, while black holes in low- to intermediate-mass galaxies remain comparatively of low mass or even absent \citep{Scott_2013_NSC, Neumayer_2020_NSCrev}.

\citet{Yao_Yuhan_2025_24tvd} placed an upper limit of $M_{\star,\rm NSC} < 10^{7.6}~M_\odot$ based on the absence of residual flux at the TDE position after performing \texttt{scarlet} scene modeling on pre-TDE ground-based DESI Legacy Imaging Survey data \citep{2018A&C....24..129M, 2019AJ....157..168D}. Their constraint was derived from the $g$-band limiting magnitude, converted to a stellar mass limit by adopting a mass-to-light ratio similar to that of the satellite dwarf galaxy associated with EP240222a \citep{Jin_2025_EP24}.

Our inferred $M_{\star,\rm NSC}$ is higher by a factor of $\sim 2$ relative to their reported limit. However, given the uncertainties in the conversion from flux to mass—particularly the sensitivity to the assumed stellar population properties\footnote{For example, if the NSC hosts an older stellar population, as is the case for the host galaxy of AT\,2024tvd compared to the satellite galaxy in EP240222a, the true mass-to-light ratio would be higher, making the \citet{Yao_Yuhan_2025_24tvd} upper limit more stringent in flux but less constraining in mass.}—we do not regard this discrepancy as a significant inconsistency. Based on the median of the posterior distributions for the NSC’s mass, age, and metallicity, we estimate an $R$-band magnitude of $\sim23$~AB~mag. Such an NSC should be readily detectable by \textit{HST} and \textit{JWST} once the TDE fades further.

\textbf{Age and Metallicity of the NSC:} 
The age and metallicity of the NSC are effectively unconstrained in our SED modeling, as the fit relies primarily on the continuum shape of the \textit{JWST} infrared spectrum. Stellar age is most strongly imprinted in absorption features and the blue/UV part of the spectrum, whereas at longer wavelengths the continuum is only weakly sensitive to age, and thus plays little role in constraining this parameter. Similarly, the metallicity parameter in our fits tends to run against the lower prior boundary ($\log Z/Z_\odot = -2$), reflecting the fact that continuum–only fitting provides limited leverage on stellar populations. In practice, meaningful constraints on the NSC age and metallicity would require inclusion of high–S/N optical or near–UV spectroscopy where metal–sensitive absorption lines are present. As such, while our modeling identifies the presence of an NSC, its detailed stellar population properties remain uncertain.

A true measurement of the NSC’s age will likely require spectroscopy, but an independent early clue may come from the detection of the Bowen N\,III~$\lambda4640$ complex in the KCWI optical spectrum. This feature indicates that the disrupted star’s debris was enriched in CNO-cycle-processed gas, suggesting it originated from a moderately massive ($\sim$1--5~$M_\odot$) star. Such a star would be younger than the $\sim12$~Gyr-old bulge population of the host galaxy, whose 0.8~$M_\odot$ main-sequence stars burn hydrogen primarily through the proton--proton chain instead. If a younger NSC stellar population is confirmed with future observations, it would be a clear indication that the offset MBH is the product of an external galaxy merger.

\textbf{Dust luminosity and temperature:} The infrared excess in the SED is well described by a single blackbody component. Our fits yield a temperature of $T_{\rm dust} = 873^{+15}_{-14}$~K and a peak luminosity of $\log L_{\rm dust}/{\rm erg~s^{-1}} = 40.80 \pm 0.01$. These values are in good agreement with dust echoes observed in other TDEs, which typically show warm dust components with $T \approx 500$–$1500$~K and luminosities of $L_{\rm dust} \approx 10^{40}$–$10^{42}$~erg~s$^{-1}$ \citep[e.g.,][]{Jiang_2021_IRechoes, Jiang_2016_dustecho, vanvelzen_2016_TDedust, vanVelzen_2021_dust, 2023ApJ...959L..19D}.  

At any time $t$, the illuminated dust lies along a paraboloid (isodelay) surface set by the light-travel time. At the epoch of our \textit{JWST} observations ($+199$~days), this geometry limits the maximum dust echo radius to $R \lesssim c\,t \simeq 200$ light days, beyond which the dust has not yet been illuminated. For consistency, we compare this with the dust sublimation radius,
\begin{equation}
R_{\rm sub} \approx 0.1~{\rm pc} \,
\left( \frac{L_{\rm UV}}{10^{45}~{\rm erg~s^{-1}}} \right)^{1/2},
\end{equation}

\citep[e.g.,][]{Netzer_2015_AGNrev}, 
which for a peak SED luminosity of $10^{44}\,{\rm erg\,s^{-1}}$, and sublimation temperature of 1800\,K yields $R_{\rm sub} \approx {\rm 35}$ light days. This value lies within the maximum echo radius, indicating that the observed infrared excess is self-consistent with reprocessing by dust at or near the sublimation front. 
Finally, the absence of CO overtone features in the infrared spectrum argues against in-situ dust formation, supporting the origin of emission from pre-existing circum-TDE dust heated to near its sublimation temperature.

\subsection{The origin of the TDE-causing black hole}

Several scenarios can, in principle, explain the presence of an apparently off–nuclear MBH. One possibility is a gravitational recoil following an MBH merger, in which anisotropic gravitational-wave emission displaces the remnant MBH from the galaxy nucleus (e.g., \citealt{Komossa_2012_recoilBHrev}). However, this mechanism is unlikely in AT\,2024tvd: the inferred black hole mass of $M_\bullet \approx 10^{5.5}\,M_\odot$ is far too small compared to the host’s central black hole mass of $\sim 10^{8.5}\,M_\odot$, making it implausible that the lighter MBH represents a recoiling remnant of a binary MBH merger. Another option is that the TDE-causing MBH was dynamically ejected from the galactic center due to multi-body interactions or nuclear asymmetries \citep{Volonteri_Perna_2005_ejectMBH, 2020_Naoz}. Yet, we find no evidence for velocity offsets between the TDE spectrum and the systemic redshift of the host, nor do we detect kinematic disturbances in the bulge. While the absence of such features cannot entirely rule out a dynamical ejection, these observations do not favor this scenario. We can, however, be confident that if the offset TDE did originate from dynamical ejection, the ejected MBH was likely of relatively low mass as opposed to a $\gtrsim 10^{8} M_{\odot}$ black hole.

The most natural explanation is instead a minor merger origin \citep{Tremmel_2018_wandMBH}. In this picture, a lower-mass satellite galaxy, hosting its own MBH and NSC, has merged with the massive host. Dynamical friction will, over time, cause the merged MBH to sink towards the host galaxy's nucleus, eventually forming a bound binary MBH system. Although the timescale of such binary formation is uncertain, observations already point to SMBH pairs across a wide range of separations---from sub-parsec/parsec-scale candidates to kiloparsec-scale dual AGN (e.g., \citealp{1988_Sillanpaa, 2006_Rodriguez, 2017_bansal, 2014_Deane, 2003_Komossa}). Such systems are likely the results of major galaxy mergers, however, the hierarchical nature of the galaxy formation suggests that minor mergers should occur much more frequently, and might even be important for driving large changes in AGN behavior \citep[e.g.,][]{Dodd2025}.

In the case of AT\,2024tvd, several pieces of evidence support the minor-merger interpretation. First, the inferred NSC mass of $M_{\star,\mathrm{NSC}} \approx 10^{8}\,M_\odot$ is consistent with a nucleated satellite of stellar mass $\sim 10^{10}\,M_\odot$, implying a merger mass ratio of order $1{:}10$, squarely in the minor–merger regime. An alternative interpretation of the compact stellar component is that it may be an ultra–compact dwarf (UCD) galaxy rather than a classical NSC. UCDs are thought to be either massive star clusters or the tidally stripped nuclei of galaxies. They typically span masses of $10^{6}$–$10^{8}\,M_\odot$ and effective radii of 10–100 pc, which is unresolved in \textit{HST} and \textit{JWST} imaging at the redshift of AT\,2024tvd \citep[e.g.,][]{Pfeffer_2013_UCD, Norris_2014_UCD}. The inferred stellar mass of $M_{\star,\mathrm{NSC}}\!\approx\!10^{8}\,M_\odot$ lies at the upper end of the UCD distribution, making a stripped-nucleus origin plausible. In this picture, AT\,2024tvd could be associated with a disrupted satellite whose compact nucleus survives as a UCD–like remnant, hosting the $10^{5.5}\,M_\odot$ black hole ultimately responsible for the TDE. A similar case has been observed for the TDE AT\,2022wtn, which occured in an interacting satellite galaxy with a merger ratio $1{:}10$ and appears to be in the early stages of a minor merger \citep{2025_Onori}. 

Second, the host morphology is remarkably smooth: the \textit{JWST} imaging shows no tidal tails or shell structures, and the KCWI and NIRSpec kinematic maps reveal no secondary kinematic components. This lack of large–scale disturbances is consistent with a minor merger, since such interactions can deposit a compact nucleus and its black hole without producing the dramatic tidal features expected from major mergers \citep[e.g.,][]{2021_Dodd, Naab_2003_galmerg, Hopkins_2010_galmerg}. 

Third, the host bulge shows strong rotational support, suggesting a system that has experienced a relatively minor interaction and is still in the process of dynamical relaxation. The bulge of the host galaxy consists of an old stellar population ($\sim$12--13 Gyr) yet shows clear rotational support in our kinematic maps. This combination is unusual in the classical bulge/pseudobulge framework: classical bulges are typically old and dispersion-dominated, whereas pseudobulges are younger and rotationally supported \citep{Kormendy_2004_bulge, Gadotti_2009_bulge}. One plausible explanation is that the system hosts a composite bulge, where an old, classical component dominates the light while a flatter, rotating pseudobulge contributes significantly to the observed kinematics \citep[e.g.,][]{Erwin_2015_bulge}. But the data are also consistent with a past minor merger, which can spin up bulges and disks without dramatically altering the old stellar population \citep{Naab_2009_minormerg}.

\section{Conclusion}
\label{sec:conclusion}

AT\,2024tvd is the first optically selected off-nuclear TDE. In this work we presented \textit{JWST} and Keck/KCWI observations aimed at understanding the origin of the off-centered MBH and exploiting the fact that, being displaced from the nucleus, the TDE suffers less contamination from nuclear starlight. This provided a comparatively cleaner environment for detailed SED analysis, including a test for whether a nuclear star cluster surrounds the offset MBH.  

Using \textit{JWST}/NIRSpec and Keck/KCWI, we extracted optical and infrared spectra, which revealed broad H and He emission lines consistent with a TDE classification. Stellar kinematics (radial velocity, velocity dispersion, age, and metallicity) were measured with \texttt{pPXF}. Both \textit{JWST} and Keck data show smooth host morphology with no secondary kinematic component or tidal features. The bulge exhibits ordered rotation, suggesting that it is still in the process of dynamical relaxation.  

We constructed a quasi-simultaneous SED of the TDE from the X-ray through the infrared at $+199$ days, and identified three emission components: (i) the TDE accretion flow, from which we constrained the black hole mass ($\sim 10^{5.5} M_{\odot}$), accretion rate, and disk outer radius; (ii) an NSC, characterized by stellar mass, age, and metallicity; and (iii) dust-echo emission, giving us the dust luminosity and temperature. The SED modeling statistically favors the presence of a massive compact stellar system with $M_{\star,\mathrm{NSC}} \approx 10^{8}\,M_\odot$ at the TDE location.  

We find that the offset MBH in AT\,2024tvd is best explained by a minor–merger scenario in which a nucleated satellite galaxy, hosting a $10^{5.5}\,M_\odot$ black hole embedded in a massive NSC, merged with the more massive host. The absence of large-scale tidal features or secondary kinematic structures is consistent with the outcome of a relatively gentle, low/intermediate–mass–ratio merger.  

Future \textit{JWST} observations, once the TDE has further faded, will enable not only a visual confirmation but also improved characterization of the NSC stellar population, including its age and metallicity. AT\,2024tvd also provides a rare opportunity to follow the long-term evolution of a TDE accretion disk in a comparatively clean off-nuclear environment. Time-domain SED measurements of this source will therefore be particularly valuable.

%% Please use the acknowledgment and contribution environments. This will 
%% be anonomyized when the "anonymous" style option is used. 
\begin{acknowledgments}

The UCSC transients team is supported in part by STcI grant JWST-DD-9249 and by a fellowship from the David and Lucile Packard Foundation to R.J.F. E.R.R.\ acknowledges the Heising-Simons Foundation and NSF: AST--1852393, AST--2150255, and AST--2206243. 

This work is based in part on observations made with the NASA/ESA/CSA James Webb Space Telescope. The data were obtained from the Mikulski Archive for Space Telescopes at the Space Telescope Science Institute, which is operated by the Association of Universities for Research in Astronomy, Inc., under NASA contract NAS 5--03127 for JWST. These observations are associated with program 9249 (PI Patra).  Support for program 9249 was provided by NASA through a grant from the Space Telescope Science Institute, which is operated by the Association of Universities for Research in Astronomy, Inc., under NASA contract NAS 5-03127.

Some of the data presented herein were obtained at Keck Observatory, which is a private 501(c)3 non-profit organization operated as a scientific partnership among the California Institute of Technology, the University of California, and the National Aeronautics and Space Administration. The Observatory was made possible by the generous financial support of the W.\ M.\ Keck Foundation.  The authors wish to recognize and acknowledge the very significant cultural role and reverence that the summit of Maunakea has always had within the Native Hawaiian community. We are most fortunate to have the opportunity to conduct observations from this mountain.

\end{acknowledgments}

\begin{contribution}
KCP and RF developed the initial research plan and led the efforts to write the proposal and acquire the data.
KCP led the data analysis and the writing of the manuscript.
NE and KDF analyzed the \textit{Swift} XRT and UVOT data and assisted with manuscript editing.
KWD led the KCWI observations and assisted with manuscript editing.
ERR contributed to the discussion section and helped edit the manuscript.
VAV, SG, KT, PA, PM, RK, and ST contributed to the proposal writing and manuscript editing.

%%This section gives authors the space to recognize author contributions. The text inside this environment is NOT counted towards the total word quanta. At a minimum, manuscripts are expected to include this text:

%All authors contributed equally to the Terra Mater collaboration.

%% But authors are expected to provide more specific details, e.g. 
%%
%%SC was responsible for writing and submitting the manuscript.
%%WWM came up with the initial research concept and edited the manuscript.
%%OTS obtained the funding and edited the manuscript.
%%EBF provided the formal analysis and validation. He also edited the manuscript.
%%GEH Supervised the undergraduates, wrote the software and administers the project github and Zenodo repositories.
%%
%% Authors can use the Contributor Role Taxonomy (CRediT) at
%% for ideas on how write a good statement tailored to their needs.

\end{contribution}

% \facilities{James Webb Space Telescope, Keck Cosmic Web Imager, \textit{Swift}, Zwicky Transient Facility}

\facilities{JWST (NIRCam, NIRSpec), Keck:II (KCWI), Swift (UVOT, XRT), PO:1.2m}

\software{\texttt{astropy} \citep{2013A&A...558A..33A,2018AJ....156..123A,2022ApJ...935..167A}, \texttt{emcee} \citep{Foreman-Mackey_etal_2013_emcee}, \texttt{pPXF} \citep{Cappellari_2012_ppxfsoft, Cappellari_2017_ppxfpaper}, \texttt{heasoft} \citep{HEASOFT}, \texttt{STPSF} \citep{Perrin_2025_STPSF}, \texttt{JWST Calibration Pipeline} \citep{Bushouse_2023_pipeline}, \texttt{Wicked} \citep{Dumont_2025_wicked}
          }


\appendix
\label{sec:appendix}

\section{Additional Figures and Tables}

\begin{table}[ht]
\centering
\caption{NIRCam Flux Measurements of AT\,2024tvd}
\begin{tabular}{lccc}
\hline\hline
Filter & Flux [$\mu$Jy] & $\sigma_{\rm stat}$ [$\mu$Jy] & $\sigma_{\rm sys}$ [$\mu$Jy] \\
\hline
F090W  & 10.81 & 0.18 & 1.62 \\
F150W  & 11.91 & 0.17 & 1.07 \\
F277W  & 14.66 & 0.12 & 0.73 \\
F444W  & 19.44 & 0.12 & 0.78 \\
\hline
\end{tabular}
\label{tab:nircam_flux}
\end{table}

\begin{table*}
\centering
\caption{Summary of model parameters fitted with \texttt{emcee}. Priors are uniform within the given ranges unless otherwise specified. Posterior values represent the median and 68\% credible intervals.}
\label{tbl:emcee_results}
\begin{tabular}{lccc}
\hline\hline
Parameter & Description & Prior Range & Posterior \\
\hline
$\log_{10} M_\star$ & NSC mass [$M_\odot$] & $3 \leq \log_{10} M_\star \leq 10$ &                                   $7.97^{+0.16}_{-0.26}$ \\
Age & NSC age [Gyr] & $1 \leq {\rm Age} \leq 15$ &                                                                  $5.1^{+5.3}_{-3.6}$ \\
$\log_{10} (Z/Z_\odot)$ & NSC metallicity & $-2.0 \leq \log_{10} Z \leq +0.5$ &                                     $-1.74^{+0.23}_{-0.18}$ \\
$\log_{10} M_{\rm BH}$ & Black hole mass [$M_\odot$] & $4 \leq \log_{10} M_{\rm BH} \leq 9$ &                       $5.50^{+0.04}_{-0.04}$ \\
$\log_{10} \dot{M}$ & Mass accretion rate [$M_\odot\,{\rm yr}^{-1}$] & $-4 \leq \log_{10} \dot{M} \leq 2$ &         $-1.22^{+0.04}_{-0.04}$ \\
$\log_{10} (r_{\rm out}/r_g)$ & Outer disk radius [$r_g$] & $1 \leq \log_{10} (r_{\rm out}/r_g) \leq 7$ &           $3.82^{+0.12}_{-0.11}$ \\
%$\log_{10} (r_p/r_g)$ & Pericenter distance [$r_g$] & $1 \leq \log_{10} (r_p/r_g) \leq 3$ (with $r_p \leq r_{\rm out}$) & $1.65^{+0.88}_{-0.94}$ \\
$\log_{10} A_{\rm peak}$ & Dust emission normalization [erg s$^{-1}$] & $40 \leq \log A_{\rm peak} \leq 43$ &       $40.80^{+0.01}_{-0.01}$ \\
$T_{\rm dust}$ & Dust temperature [K] & $100 \leq T_{\rm dust} \leq 2000$ &                                         $873^{+15}_{-14}$ \\
$\ln s_{0}$ & Intrinsic scatter [dex] & $-10 \leq \ln s_{0} \leq 0$ &                                               $-6.5^{+2.0}_{-2.4}$ \\
\hline
\end{tabular}
\end{table*}

\begin{figure*}[ht!]
    \centering
    \includegraphics[width=\textwidth]{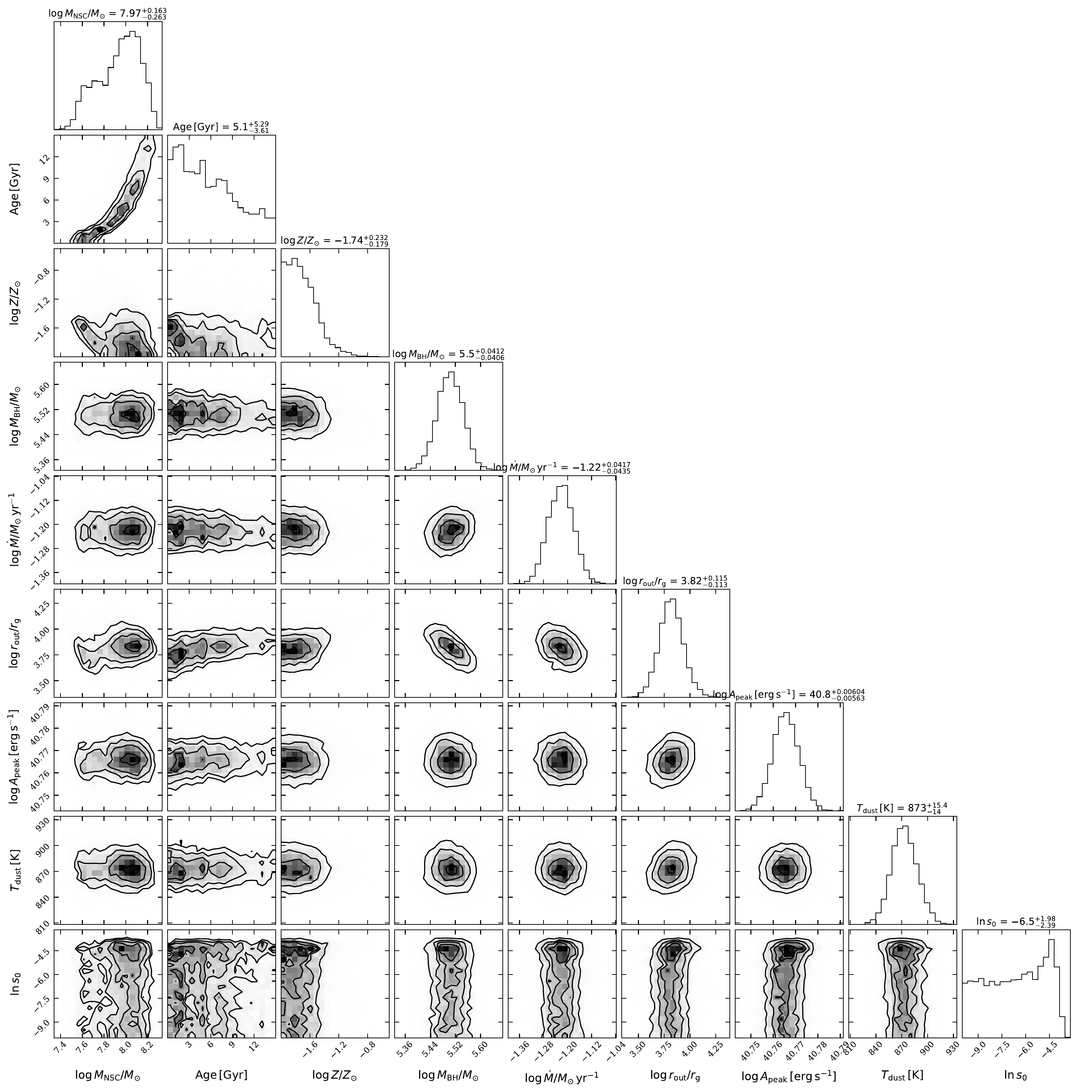}
    \caption{Posterior distributions of the fitted model parameters when the SED includes an NSC component.}
    \label{fig:corner_NSC}
\end{figure*}

\begin{figure*}[ht!]
    \centering
    \includegraphics[width=\textwidth]{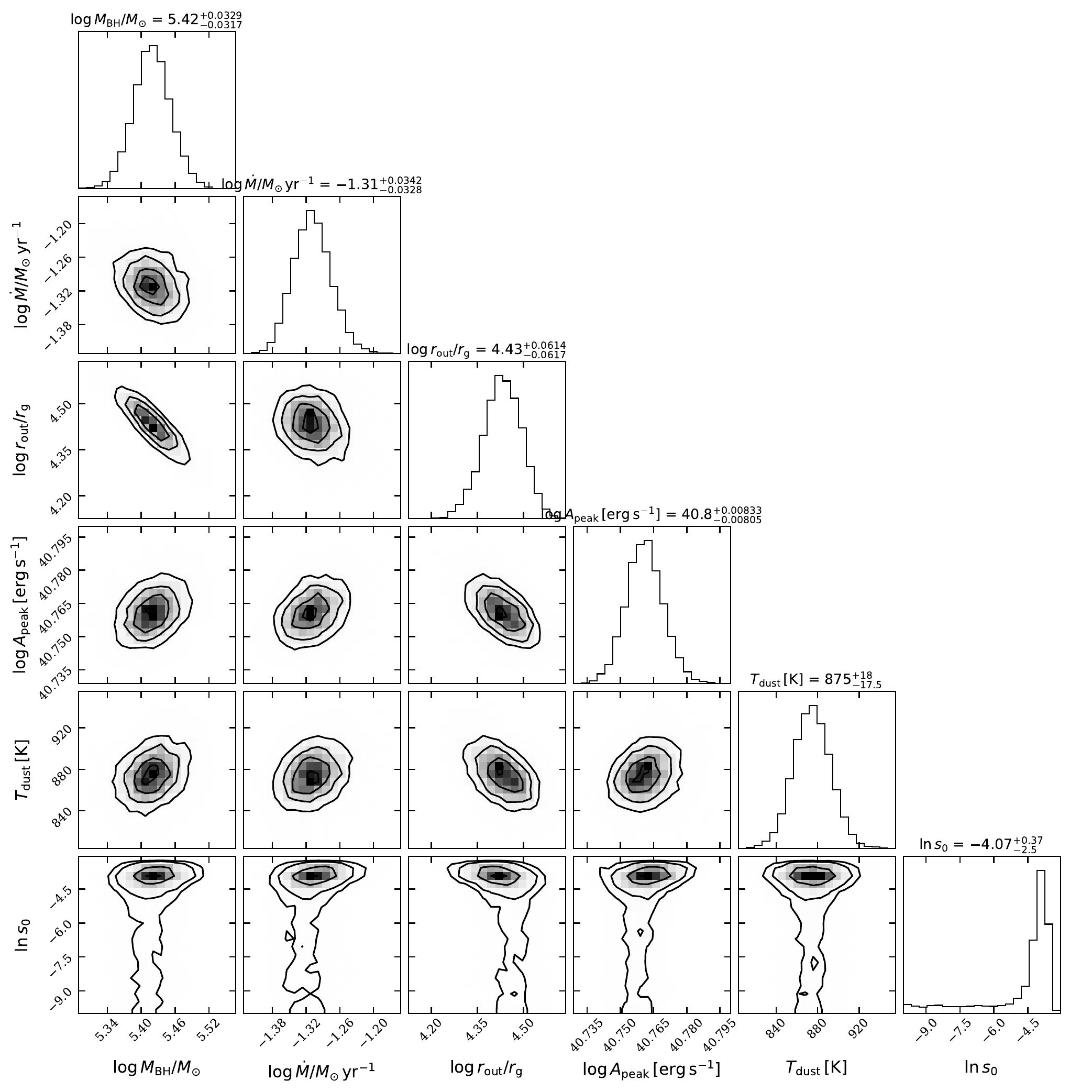}
    \caption{Posterior distributions of the fitted model parameters without an NSC component.}
    \label{fig:corner_noNSC}
\end{figure*}

\begin{figure*}[ht!]
    \centering
    % ---- Top row ----
    \includegraphics[width=0.48\textwidth]{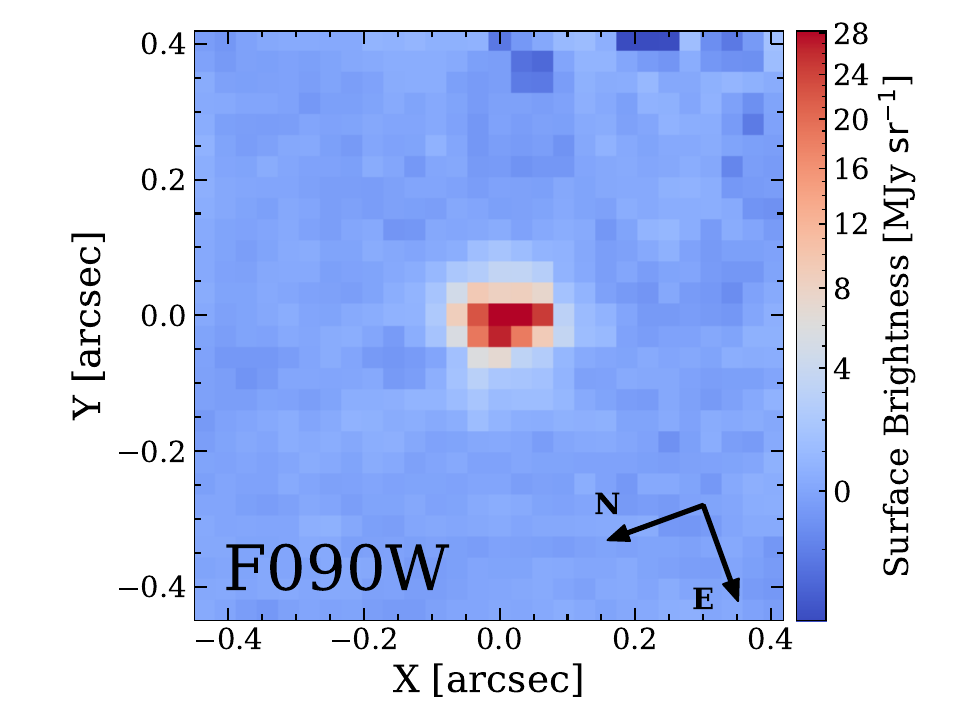}\hfill
    \includegraphics[width=0.48\textwidth]{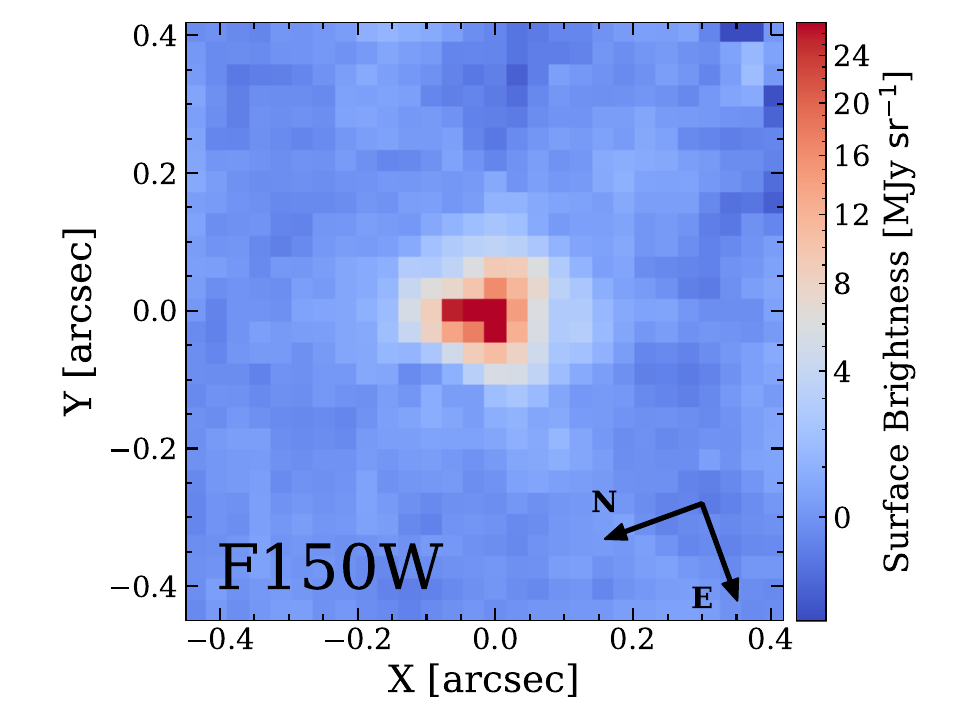}\\[1ex]
    % ---- Bottom row ----
    \includegraphics[width=0.48\textwidth]{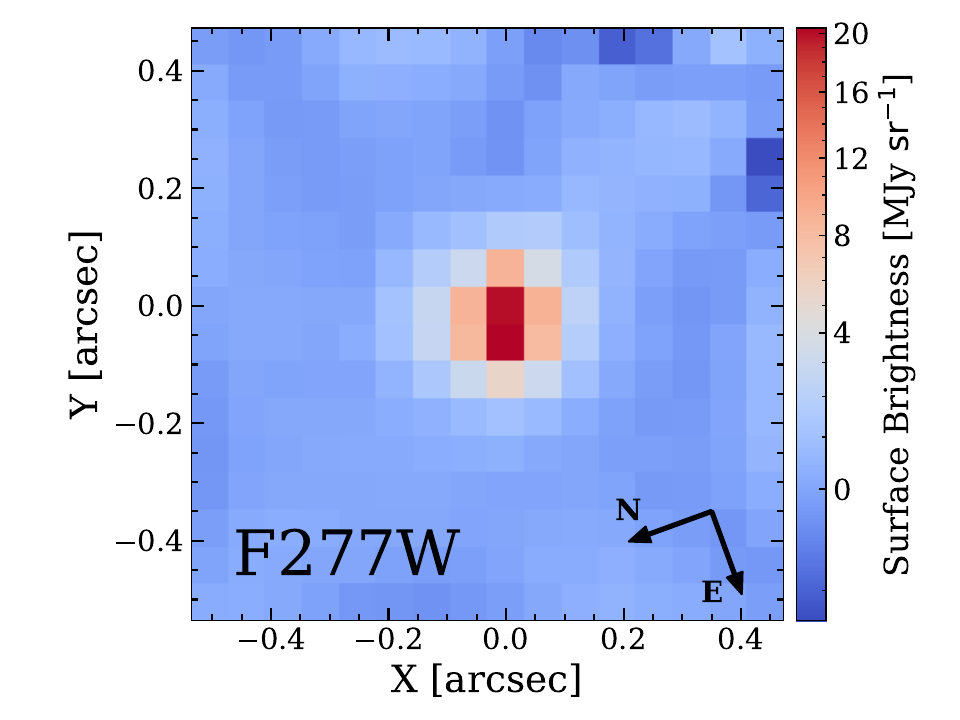}\hfill
    \includegraphics[width=0.48\textwidth]{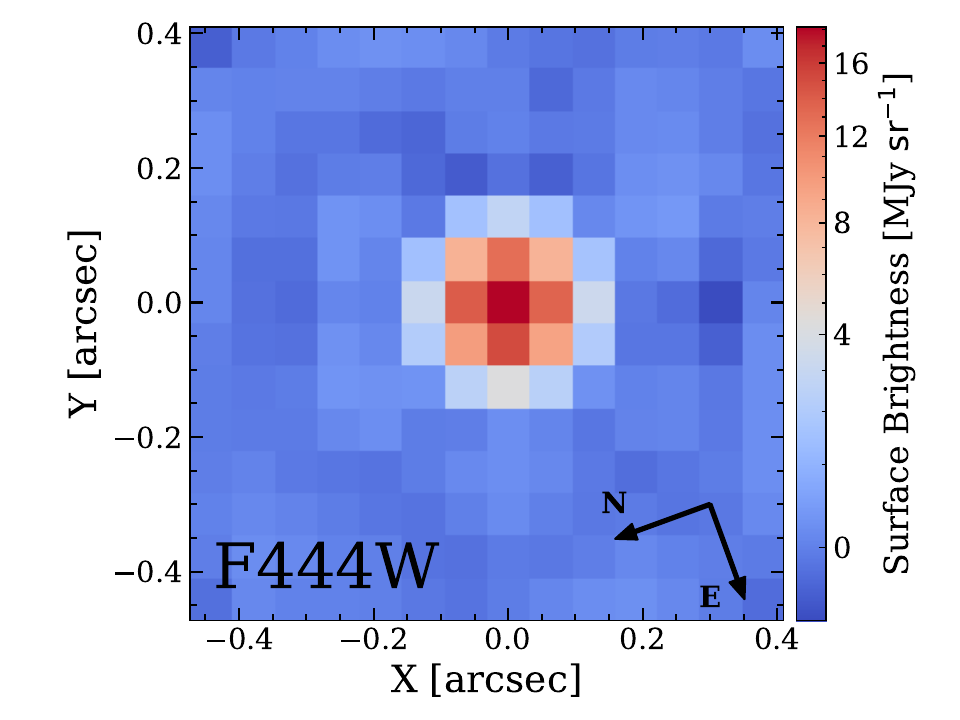}
    \caption{NIRCam image cutouts centered on AT\,2024tvd. The underlying host-galaxy light was modeled with a two-dimensional polynomial surface and subtracted. The host nucleus is towards the upper right corner. The residual images show no significant structure other than the PSF wings.}
    \label{fig:nircam_bkg_sub}
\end{figure*}

\bibliography{references}{}

\begin{thebibliography}{}
\expandafter\ifx\csname natexlab\endcsname\relax\def\natexlab#1{#1}\fi
\providecommand{\url}[1]{\href{#1}{#1}}
\providecommand{\dodoi}[1]{doi:~\href{http://doi.org/#1}{\nolinkurl{#1}}}
\providecommand{\doeprint}[1]{\href{http://ascl.net/#1}{\nolinkurl{http://ascl.net/#1}}}
\providecommand{\doarXiv}[1]{\href{https://arxiv.org/abs/#1}{\nolinkurl{https://arxiv.org/abs/#1}}}

\bibitem[{R. {Ahumada} {et~al.}(2020){Ahumada}, {Allende Prieto}, {Almeida},
  {Anders}, {Anderson}, {Andrews}, {Anguiano}, {Arcodia}, {Armengaud},
  {Aubert}, {Avila}, {Avila-Reese}, {Badenes}, {Balland}, {Barger},
  {Barrera-Ballesteros}, {Basu}, {Bautista}, {Beaton}, {Beers}, {Benavides},
  {Bender}, {Bernardi}, {Bershady}, {Beutler}, {Bidin}, {Bird}, {Bizyaev},
  {Blanc}, {Blanton}, {Boquien}, {Borissova}, {Bovy}, {Brandt}, {Brinkmann},
  {Brownstein}, {Bundy}, {Bureau}, {Burgasser}, {Burtin}, {Cano-D{\'\i}az},
  {Capasso}, {Cappellari}, {Carrera}, {Chabanier}, {Chaplin}, {Chapman},
  {Cherinka}, {Chiappini}, {Doohyun Choi}, {Chojnowski}, {Chung}, {Clerc},
  {Coffey}, {Comerford}, {Comparat}, {da Costa}, {Cousinou}, {Covey}, {Crane},
  {Cunha}, {Ilha}, {Dai}, {Damsted}, {Darling}, {Davidson}, {Davies}, {Dawson},
  {De}, {de la Macorra}, {De Lee}, {Queiroz}, {Deconto Machado}, {de la Torre},
  {Dell'Agli}, {du Mas des Bourboux}, {Diamond-Stanic}, {Dillon}, {Donor},
  {Drory}, {Duckworth}, {Dwelly}, {Ebelke}, {Eftekharzadeh}, {Davis Eigenbrot},
  {Elsworth}, {Eracleous}, {Erfanianfar}, {Escoffier}, {Fan}, {Farr},
  {Fern{\'a}ndez-Trincado}, {Feuillet}, {Finoguenov}, {Fofie},
  {Fraser-McKelvie}, {Frinchaboy}, {Fromenteau}, {Fu}, {Galbany}, {Garcia},
  {Garc{\'\i}a-Hern{\'a}ndez}, {Garma Oehmichen}, {Ge}, {Geimba Maia},
  {Geisler}, {Gelfand}, {Goddy}, {Gonzalez-Perez}, {Grabowski}, {Green},
  {Grier}, {Guo}, {Guy}, {Harding}, {Hasselquist}, {Hawken}, {Hayes}, {Hearty},
  {Hekker}, {Hogg}, {Holtzman}, {Horta}, {Hou}, {Hsieh}, {Huber}, {Hunt}, {Ider
  Chitham}, {Imig}, {Jaber}, {Jimenez Angel}, {Johnson}, {Jones},
  {J{\"o}nsson}, {Jullo}, {Kim}, {Kinemuchi}, {Kirkpatrick}, {Kite}, {Klaene},
  {Kneib}, {Kollmeier}, {Kong}, {Kounkel}, {Krishnarao}, {Lacerna}, {Lan},
  {Lane}, {Law}, {Le Goff}, {Leung}, {Lewis}, {Li}, {Lian}, {Lin}, {Long},
  {Longa-Pe{\~n}a}, {Lundgren}, {Lyke}, {Mackereth}, {MacLeod}, {Majewski},
  {Manchado}, {Maraston}, {Martini}, {Masseron}, {Masters}, {Mathur},
  {McDermid}, {Merloni}, {Merrifield}, {M{\'e}sz{\'a}ros}, {Miglio}, {Minniti},
  {Minsley}, {Miyaji}, {Mohammad}, {Mosser}, {Mueller}, {Muna},
  {Mu{\~n}oz-Guti{\'e}rrez}, {Myers}, {Nadathur}, {Nair}, {Nandra}, {Correa do
  Nascimento}, {Nevin}, {Newman}, {Nidever}, {Nitschelm}, {Noterdaeme},
  {O'Connell}, {Olmstead}, {Oravetz}, {Oravetz}, {Osorio}, {Pace}, {Padilla},
  {Palanque-Delabrouille}, \& {Palicio}}]{Ahumada_2020_SDSS}
{Ahumada}, R., {Allende Prieto}, C., {Almeida}, A., {et~al.} 2020,
  \bibinfo{title}{{The 16th Data Release of the Sloan Digital Sky Surveys:
  First Release from the APOGEE-2 Southern Survey and Full Release of eBOSS
  Spectra},} \apjs, 249, 3, \dodoi{10.3847/1538-4365/ab929e}

\bibitem[{ {Astropy Collaboration} {et~al.}(2013){Astropy Collaboration},
  {Robitaille}, {Tollerud}, {Greenfield}, {Droettboom}, {Bray}, {Aldcroft},
  {Davis}, {Ginsburg}, {Price-Whelan}, {Kerzendorf}, {Conley}, {Crighton},
  {Barbary}, {Muna}, {Ferguson}, {Grollier}, {Parikh}, {Nair}, {Unther},
  {Deil}, {Woillez}, {Conseil}, {Kramer}, {Turner}, {Singer}, {Fox}, {Weaver},
  {Zabalza}, {Edwards}, {Azalee Bostroem}, {Burke}, {Casey}, {Crawford},
  {Dencheva}, {Ely}, {Jenness}, {Labrie}, {Lim}, {Pierfederici}, {Pontzen},
  {Ptak}, {Refsdal}, {Servillat}, \& {Streicher}}]{2013A&A...558A..33A}
{Astropy Collaboration}, {Robitaille}, T.~P., {Tollerud}, E.~J., {et~al.} 2013,
  \bibinfo{title}{{Astropy: A community Python package for astronomy},} \aap,
  558, A33, \dodoi{10.1051/0004-6361/201322068}

\bibitem[{ {Astropy Collaboration} {et~al.}(2018){Astropy Collaboration},
  {Price-Whelan}, {Sip{\H{o}}cz}, {G{\"u}nther}, {Lim}, {Crawford}, {Conseil},
  {Shupe}, {Craig}, {Dencheva}, {Ginsburg}, {VanderPlas}, {Bradley},
  {P{\'e}rez-Su{\'a}rez}, {de Val-Borro}, {Aldcroft}, {Cruz}, {Robitaille},
  {Tollerud}, {Ardelean}, {Babej}, {Bach}, {Bachetti}, {Bakanov}, {Bamford},
  {Barentsen}, {Barmby}, {Baumbach}, {Berry}, {Biscani}, {Boquien}, {Bostroem},
  {Bouma}, {Brammer}, {Bray}, {Breytenbach}, {Buddelmeijer}, {Burke},
  {Calderone}, {Cano Rodr{\'\i}guez}, {Cara}, {Cardoso}, {Cheedella}, {Copin},
  {Corrales}, {Crichton}, {D'Avella}, {Deil}, {Depagne}, {Dietrich}, {Donath},
  {Droettboom}, {Earl}, {Erben}, {Fabbro}, {Ferreira}, {Finethy}, {Fox},
  {Garrison}, {Gibbons}, {Goldstein}, {Gommers}, {Greco}, {Greenfield},
  {Groener}, {Grollier}, {Hagen}, {Hirst}, {Homeier}, {Horton}, {Hosseinzadeh},
  {Hu}, {Hunkeler}, {Ivezi{\'c}}, {Jain}, {Jenness}, {Kanarek}, {Kendrew},
  {Kern}, {Kerzendorf}, {Khvalko}, {King}, {Kirkby}, {Kulkarni}, {Kumar},
  {Lee}, {Lenz}, {Littlefair}, {Ma}, {Macleod}, {Mastropietro}, {McCully},
  {Montagnac}, {Morris}, {Mueller}, {Mumford}, {Muna}, {Murphy}, {Nelson},
  {Nguyen}, {Ninan}, {N{\"o}the}, {Ogaz}, {Oh}, {Parejko}, {Parley}, {Pascual},
  {Patil}, {Patil}, {Plunkett}, {Prochaska}, {Rastogi}, {Reddy Janga},
  {Sabater}, {Sakurikar}, {Seifert}, {Sherbert}, {Sherwood-Taylor}, {Shih},
  {Sick}, {Silbiger}, {Singanamalla}, {Singer}, {Sladen}, {Sooley},
  {Sornarajah}, {Streicher}, {Teuben}, {Thomas}, {Tremblay}, {Turner},
  {Terr{\'o}n}, {van Kerkwijk}, {de la Vega}, {Watkins}, {Weaver}, {Whitmore},
  {Woillez}, {Zabalza}, \& {Astropy Contributors}}]{2018AJ....156..123A}
{Astropy Collaboration}, {Price-Whelan}, A.~M., {Sip{\H{o}}cz}, B.~M., {et~al.}
  2018, \bibinfo{title}{{The Astropy Project: Building an Open-science Project
  and Status of the v2.0 Core Package},} \aj, 156, 123,
  \dodoi{10.3847/1538-3881/aabc4f}

\bibitem[{ {Astropy Collaboration} {et~al.}(2022{\natexlab{a}}){Astropy
  Collaboration}, {Price-Whelan}, {Lim}, {Earl}, {Starkman}, {Bradley},
  {Shupe}, {Patil}, {Corrales}, {Brasseur}, {N{\"o}the}, {Donath}, {Tollerud},
  {Morris}, {Ginsburg}, {Vaher}, {Weaver}, {Tocknell}, {Jamieson}, {van
  Kerkwijk}, {Robitaille}, {Merry}, {Bachetti}, {G{\"u}nther}, {Aldcroft},
  {Alvarado-Montes}, {Archibald}, {B{\'o}di}, {Bapat}, {Barentsen},
  {Baz{\'a}n}, {Biswas}, {Boquien}, {Burke}, {Cara}, {Cara}, {Conroy},
  {Conseil}, {Craig}, {Cross}, {Cruz}, {D'Eugenio}, {Dencheva}, {Devillepoix},
  {Dietrich}, {Eigenbrot}, {Erben}, {Ferreira}, {Foreman-Mackey}, {Fox},
  {Freij}, {Garg}, {Geda}, {Glattly}, {Gondhalekar}, {Gordon}, {Grant},
  {Greenfield}, {Groener}, {Guest}, {Gurovich}, {Handberg}, {Hart},
  {Hatfield-Dodds}, {Homeier}, {Hosseinzadeh}, {Jenness}, {Jones}, {Joseph},
  {Kalmbach}, {Karamehmetoglu}, {Ka{\l}uszy{\'n}ski}, {Kelley}, {Kern},
  {Kerzendorf}, {Koch}, {Kulumani}, {Lee}, {Ly}, {Ma}, {MacBride}, {Maljaars},
  {Muna}, {Murphy}, {Norman}, {O'Steen}, {Oman}, {Pacifici}, {Pascual},
  {Pascual-Granado}, {Patil}, {Perren}, {Pickering}, {Rastogi}, {Roulston},
  {Ryan}, {Rykoff}, {Sabater}, {Sakurikar}, {Salgado}, {Sanghi}, {Saunders},
  {Savchenko}, {Schwardt}, {Seifert-Eckert}, {Shih}, {Jain}, {Shukla}, {Sick},
  {Simpson}, {Singanamalla}, {Singer}, {Singhal}, {Sinha}, {Sip{\H{o}}cz},
  {Spitler}, {Stansby}, {Streicher}, {{\v{S}}umak}, {Swinbank}, {Taranu},
  {Tewary}, {Tremblay}, {de Val-Borro}, {Van Kooten}, {Vasovi{\'c}}, {Verma},
  {de Miranda Cardoso}, {Williams}, {Wilson}, {Winkel}, {Wood-Vasey}, {Xue},
  {Yoachim}, {Zhang}, {Zonca}, \& {Astropy Project
  Contributors}}]{Astropy_2022}
{Astropy Collaboration}, {Price-Whelan}, A.~M., {Lim}, P.~L., {et~al.}
  2022{\natexlab{a}}, \bibinfo{title}{{The Astropy Project: Sustaining and
  Growing a Community-oriented Open-source Project and the Latest Major Release
  (v5.0) of the Core Package},} \apj, 935, 167,
  \dodoi{10.3847/1538-4357/ac7c74}

\bibitem[{ {Astropy Collaboration} {et~al.}(2022{\natexlab{b}}){Astropy
  Collaboration}, {Price-Whelan}, {Lim}, {Earl}, {Starkman}, {Bradley},
  {Shupe}, {Patil}, {Corrales}, {Brasseur}, {N{\"o}the}, {Donath}, {Tollerud},
  {Morris}, {Ginsburg}, {Vaher}, {Weaver}, {Tocknell}, {Jamieson}, {van
  Kerkwijk}, {Robitaille}, {Merry}, {Bachetti}, {G{\"u}nther}, {Aldcroft},
  {Alvarado-Montes}, {Archibald}, {B{\'o}di}, {Bapat}, {Barentsen},
  {Baz{\'a}n}, {Biswas}, {Boquien}, {Burke}, {Cara}, {Cara}, {Conroy},
  {Conseil}, {Craig}, {Cross}, {Cruz}, {D'Eugenio}, {Dencheva}, {Devillepoix},
  {Dietrich}, {Eigenbrot}, {Erben}, {Ferreira}, {Foreman-Mackey}, {Fox},
  {Freij}, {Garg}, {Geda}, {Glattly}, {Gondhalekar}, {Gordon}, {Grant},
  {Greenfield}, {Groener}, {Guest}, {Gurovich}, {Handberg}, {Hart},
  {Hatfield-Dodds}, {Homeier}, {Hosseinzadeh}, {Jenness}, {Jones}, {Joseph},
  {Kalmbach}, {Karamehmetoglu}, {Ka{\l}uszy{\'n}ski}, {Kelley}, {Kern},
  {Kerzendorf}, {Koch}, {Kulumani}, {Lee}, {Ly}, {Ma}, {MacBride}, {Maljaars},
  {Muna}, {Murphy}, {Norman}, {O'Steen}, {Oman}, {Pacifici}, {Pascual},
  {Pascual-Granado}, {Patil}, {Perren}, {Pickering}, {Rastogi}, {Roulston},
  {Ryan}, {Rykoff}, {Sabater}, {Sakurikar}, {Salgado}, {Sanghi}, {Saunders},
  {Savchenko}, {Schwardt}, {Seifert-Eckert}, {Shih}, {Jain}, {Shukla}, {Sick},
  {Simpson}, {Singanamalla}, {Singer}, {Singhal}, {Sinha}, {Sip{\H{o}}cz},
  {Spitler}, {Stansby}, {Streicher}, {{\v{S}}umak}, {Swinbank}, {Taranu},
  {Tewary}, {Tremblay}, {de Val-Borro}, {Van Kooten}, {Vasovi{\'c}}, {Verma},
  {de Miranda Cardoso}, {Williams}, {Wilson}, {Winkel}, {Wood-Vasey}, {Xue},
  {Yoachim}, {Zhang}, {Zonca}, \& {Astropy Project
  Contributors}}]{2022ApJ...935..167A}
{Astropy Collaboration}, {Price-Whelan}, A.~M., {Lim}, P.~L., {et~al.}
  2022{\natexlab{b}}, \bibinfo{title}{{The Astropy Project: Sustaining and
  Growing a Community-oriented Open-source Project and the Latest Major Release
  (v5.0) of the Core Package},} \apj, 935, 167,
  \dodoi{10.3847/1538-4357/ac7c74}

\bibitem[{F. {Auch{\`e}re} {et~al.}(2023){Auch{\`e}re}, {Berghmans},
  {Dumesnil}, {Halain}, {Mercier}, {Rochus}, {Delmotte}, {Fran{\c{c}}ois},
  {Hermans}, {Hervier}, {Kraaikamp}, {Meltchakov}, {Morinaud}, {Philippon},
  {Smith}, {Stegen}, {Verbeeck}, {Zhang}, {Andretta}, {Abbo}, {Buchlin},
  {Frassati}, {Gissot}, {Gyo}, {Harra}, {Jerse}, {Landini}, {Mierla}, {Nicula},
  {Parenti}, {Renotte}, {Romoli}, {Russano}, {Sasso}, {Sch{\"u}hle}, {Schmutz},
  {Soubri{\'e}}, {Susino}, {Teriaca}, {West}, \&
  {Zhukov}}]{Auchere_2023_unsharp}
{Auch{\`e}re}, F., {Berghmans}, D., {Dumesnil}, C., {et~al.} 2023,
  \bibinfo{title}{{Beyond the disk: EUV coronagraphic observations of the
  Extreme Ultraviolet Imager on board Solar Orbiter},} \aap, 674, A127,
  \dodoi{10.1051/0004-6361/202346039}

\bibitem[{K. {Auchettl} {et~al.}(2017){Auchettl}, {Guillochon}, \&
  {Ramirez-Ruiz}}]{2017_Auchettl}
{Auchettl}, K., {Guillochon}, J., \& {Ramirez-Ruiz}, E. 2017,
  \bibinfo{title}{{New Physical Insights about Tidal Disruption Events from a
  Comprehensive Observational Inventory at X-Ray Wavelengths},} \apj, 838, 149,
  \dodoi{10.3847/1538-4357/aa633b}

\bibitem[{K. {Bansal} {et~al.}(2017){Bansal}, {Taylor}, {Peck}, {Zavala}, \&
  {Romani}}]{2017_bansal}
{Bansal}, K., {Taylor}, G.~B., {Peck}, A.~B., {Zavala}, R.~T., \& {Romani},
  R.~W. 2017, \bibinfo{title}{{Constraining the Orbit of the Supermassive Black
  Hole Binary 0402+379},} \apj, 843, 14, \dodoi{10.3847/1538-4357/aa74e1}

\bibitem[{J.~E. {Barnes} \& L. {Hernquist}(1992){Barnes} \&
  {Hernquist}}]{Barnes_1992_galmergrev}
{Barnes}, J.~E., \& {Hernquist}, L. 1992, \bibinfo{title}{{Dynamics of
  interacting galaxies.},} \araa, 30, 705,
  \dodoi{10.1146/annurev.aa.30.090192.003421}

\bibitem[{E.~C. {Bellm} {et~al.}(2019){Bellm}, {Kulkarni}, {Graham}, {Dekany},
  {Smith}, {Riddle}, {Masci}, {Helou}, {Prince}, {Adams}, {Barbarino},
  {Barlow}, {Bauer}, {Beck}, {Belicki}, {Biswas}, {Blagorodnova}, {Bodewits},
  {Bolin}, {Brinnel}, {Brooke}, {Bue}, {Bulla}, {Burruss}, {Cenko}, {Chang},
  {Connolly}, {Coughlin}, {Cromer}, {Cunningham}, {De}, {Delacroix}, {Desai},
  {Duev}, {Eadie}, {Farnham}, {Feeney}, {Feindt}, {Flynn}, {Franckowiak},
  {Frederick}, {Fremling}, {Gal-Yam}, {Gezari}, {Giomi}, {Goldstein},
  {Golkhou}, {Goobar}, {Groom}, {Hacopians}, {Hale}, {Henning}, {Ho}, {Hover},
  {Howell}, {Hung}, {Huppenkothen}, {Imel}, {Ip}, {Ivezi{\'c}}, {Jackson},
  {Jones}, {Juric}, {Kasliwal}, {Kaspi}, {Kaye}, {Kelley}, {Kowalski},
  {Kramer}, {Kupfer}, {Landry}, {Laher}, {Lee}, {Lin}, {Lin}, {Lunnan},
  {Giomi}, {Mahabal}, {Mao}, {Miller}, {Monkewitz}, {Murphy}, {Ngeow},
  {Nordin}, {Nugent}, {Ofek}, {Patterson}, {Penprase}, {Porter}, {Rauch},
  {Rebbapragada}, {Reiley}, {Rigault}, {Rodriguez}, {van Roestel}, {Rusholme},
  {van Santen}, {Schulze}, {Shupe}, {Singer}, {Soumagnac}, {Stein}, {Surace},
  {Sollerman}, {Szkody}, {Taddia}, {Terek}, {Van Sistine}, {van Velzen},
  {Vestrand}, {Walters}, {Ward}, {Ye}, {Yu}, {Yan}, \&
  {Zolkower}}]{Bellm_etal_2019}
{Bellm}, E.~C., {Kulkarni}, S.~R., {Graham}, M.~J., {et~al.} 2019,
  \bibinfo{title}{{The Zwicky Transient Facility: System Overview, Performance,
  and First Results},} PASP, 131, 018002, \dodoi{10.1088/1538-3873/aaecbe}

\bibitem[{L. {Blecha} {et~al.}(2016){Blecha}, {Sijacki}, {Kelley}, {Torrey},
  {Vogelsberger}, {Nelson}, {Springel}, {Snyder}, \&
  {Hernquist}}]{2016MNRAS.456..961B}
{Blecha}, L., {Sijacki}, D., {Kelley}, L.~Z., {et~al.} 2016,
  \bibinfo{title}{{Recoiling black holes: prospects for detection and
  implications of spin alignment},} \mnras, 456, 961,
  \dodoi{10.1093/mnras/stv2646}

\bibitem[{T. {B{\"o}ker} {et~al.}(2004){B{\"o}ker}, {Sarzi}, {McLaughlin}, {van
  der Marel}, {Rix}, {Ho}, \& {Shields}}]{Boker_2004_NSC}
{B{\"o}ker}, T., {Sarzi}, M., {McLaughlin}, D.~E., {et~al.} 2004,
  \bibinfo{title}{{A Hubble Space Telescope Census of Nuclear Star Clusters in
  Late-Type Spiral Galaxies. II. Cluster Sizes and Structural Parameter
  Correlations},} \aj, 127, 105, \dodoi{10.1086/380231}

\bibitem[{H. {Bushouse} {et~al.}(2023){Bushouse}, {Eisenhamer}, {Dencheva},
  {Davies}, {Greenfield}, {Morrison}, {Hodge}, {Simon}, {Grumm}, {Droettboom},
  {Slavich}, {Sosey}, {Pauly}, {Miller}, {Jedrzejewski}, {Hack}, {Davis},
  {Crawford}, {Law}, {Gordon}, {Regan}, {Cara}, {MacDonald}, {Bradley},
  {Shanahan}, {Jamieson}, {Teodoro}, {Williams}, \&
  {Pena-Guerrero}}]{Bushouse_2023_pipeline}
{Bushouse}, H., {Eisenhamer}, J., {Dencheva}, N., {et~al.} 2023,
  \bibinfo{title}{{JWST Calibration Pipeline},}, 1.12.5 Zenodo,
  \dodoi{10.5281/zenodo.10022973}

\bibitem[{M. {Cappellari}(2012){Cappellari}}]{Cappellari_2012_ppxfsoft}
{Cappellari}, M. 2012, \bibinfo{title}{{pPXF: Penalized Pixel-Fitting stellar
  kinematics extraction},}, Astrophysics Source Code Library, record
  ascl:1210.002

\bibitem[{M. {Cappellari}(2017){Cappellari}}]{Cappellari_2017_ppxfpaper}
{Cappellari}, M. 2017, \bibinfo{title}{{Improving the full spectrum fitting
  method: accurate convolution with Gauss-Hermite functions},} \mnras, 466,
  798, \dodoi{10.1093/mnras/stw3020}

\bibitem[{M. {Cappellari} \& Y. {Copin}(2003){Cappellari} \&
  {Copin}}]{Cappellari_2003_voronoi}
{Cappellari}, M., \& {Copin}, Y. 2003, \bibinfo{title}{{Adaptive spatial
  binning of integral-field spectroscopic data using Voronoi tessellations},}
  \mnras, 342, 345, \dodoi{10.1046/j.1365-8711.2003.06541.x}

\bibitem[{J.~A. {Cardelli} {et~al.}(1989){Cardelli}, {Clayton}, \&
  {Mathis}}]{Cardelli_etal_1989}
{Cardelli}, J.~A., {Clayton}, G.~C., \& {Mathis}, J.~S. 1989,
  \bibinfo{title}{{The relationship between infrared, optical, and ultraviolet
  extinction},} ApJ, 345, 245, \dodoi{10.1086/167900}

\bibitem[{C.~T. {Christy} {et~al.}(2025){Christy}, {Alexander}, {Laskar},
  {Franz}, {Goodwin}, {Pearson}, {Berger}, {Cendes}, {Chornock}, {Coppejans},
  {Eftekhari}, {Margutti}, {Miller-Jones}, {Krips}, {Ramirez-Ruiz}, {Sand},
  {Saxton}, {Shrestha}, \& {van Velzen}}]{2025arXiv250914317C}
{Christy}, C.~T., {Alexander}, K.~D., {Laskar}, T., {et~al.} 2025,
  \bibinfo{title}{{Dichotomy in Long-Lived Radio Emission from Tidal Disruption
  Events AT 2020zso and AT 2021sdu: Multi-Component Outflows vs. Host
  Contamination},} arXiv e-prints, arXiv:2509.14317,
  \dodoi{10.48550/arXiv.2509.14317}

\bibitem[{C. {Conroy} \& J.~E. {Gunn}(2010){Conroy} \&
  {Gunn}}]{Conroy_Gunn_2010_FSPS}
{Conroy}, C., \& {Gunn}, J.~E. 2010, \bibinfo{title}{{The Propagation of
  Uncertainties in Stellar Population Synthesis Modeling. III. Model
  Calibration, Comparison, and Evaluation},} \apj, 712, 833,
  \dodoi{10.1088/0004-637X/712/2/833}

\bibitem[{C. {Conroy} {et~al.}(2009){Conroy}, {Gunn}, \&
  {White}}]{Conroy_etal_2009_FSPS}
{Conroy}, C., {Gunn}, J.~E., \& {White}, M. 2009, \bibinfo{title}{{The
  Propagation of Uncertainties in Stellar Population Synthesis Modeling. I. The
  Relevance of Uncertain Aspects of Stellar Evolution and the Initial Mass
  Function to the Derived Physical Properties of Galaxies},} \apj, 699, 486,
  \dodoi{10.1088/0004-637X/699/1/486}

\bibitem[{C.~T. {Cunningham}(1975){Cunningham}}]{Cunningham_1975_accretion}
{Cunningham}, C.~T. 1975, \bibinfo{title}{{The effects of redshifts and
  focusing on the spectrum of an accretion disk around a Kerr black hole.},}
  \apj, 202, 788, \dodoi{10.1086/154033}

\bibitem[{L. {Dai} {et~al.}(2018){Dai}, {McKinney}, {Roth}, {Ramirez-Ruiz}, \&
  {Miller}}]{Dai_etal_2018}
{Dai}, L., {McKinney}, J.~C., {Roth}, N., {Ramirez-Ruiz}, E., \& {Miller},
  M.~C. 2018, \bibinfo{title}{{A Unified Model for Tidal Disruption Events},}
  \apjl, 859, L20, \dodoi{10.3847/2041-8213/aab429}

\bibitem[{F. {De Colle} {et~al.}(2012){De Colle}, {Guillochon}, {Naiman}, \&
  {Ramirez-Ruiz}}]{2012_DeColle}
{De Colle}, F., {Guillochon}, J., {Naiman}, J., \& {Ramirez-Ruiz}, E. 2012,
  \bibinfo{title}{{The Dynamics, Appearance, and Demographics of Relativistic
  Jets Triggered by Tidal Disruption of Stars in Quiescent Supermassive Black
  Holes},} \apj, 760, 103, \dodoi{10.1088/0004-637X/760/2/103}

\bibitem[{R.~P. {Deane} {et~al.}(2014){Deane}, {Paragi}, {Jarvis}, {Coriat},
  {Bernardi}, {Fender}, {Frey}, {Heywood}, {Kl{\"o}ckner}, {Grainge}, \&
  {Rumsey}}]{2014_Deane}
{Deane}, R.~P., {Paragi}, Z., {Jarvis}, M.~J., {et~al.} 2014,
  \bibinfo{title}{{A close-pair binary in a distant triple supermassive black
  hole system},} \nat, 511, 57, \dodoi{10.1038/nature13454}

\bibitem[{J.~M. {DerKacy} {et~al.}(2025){DerKacy}, {Ashall}, {Baron}, {Medler},
  {Mera}, {Hoeflich}, {Shahbandeh}, {Burns}, {Stritzinger}, {Tucker},
  {Shappee}, {Auchettl}, {Angus}, {Desai}, {Do}, {Hinkle}, {Hoogendam},
  {Huber}, {Payne}, {Jones}, {Shi}, {Kong}, {Romagnoli}, {Syncatto}, {Moran},
  {Fereidouni}, {Brown}, {Engesser}, {Fox}, {Galbany}, {Hsiao}, {de Jaeger},
  {Kumar}, {Lu}, {Matsuura}, {Mazzali}, {Morrell}, {Pfeffer}, {Phillips},
  {Rest}, {Shiber}, {Strolger}, {Suntzeff}, {Temim}, {Tinyanont}, {Wang},
  {Wesson}, {Park}, \& {Rho}}]{Derkacy_2025_23ixf}
{DerKacy}, J.~M., {Ashall}, C., {Baron}, E., {et~al.} 2025,
  \bibinfo{title}{{JWST Observations of SN 2023ixf I: Completing the Early
  Multi-Wavelength Picture with Plateau-phase Spectroscopy},} arXiv e-prints,
  arXiv:2507.18785, \dodoi{10.48550/arXiv.2507.18785}

\bibitem[{A. {Dey} {et~al.}(2019){Dey}, {Schlegel}, {Lang}, {Blum}, {Burleigh},
  {Fan}, {Findlay}, {Finkbeiner}, {Herrera}, {Juneau}, {Landriau}, {Levi},
  {McGreer}, {Meisner}, {Myers}, {Moustakas}, {Nugent}, {Patej}, {Schlafly},
  {Walker}, {Valdes}, {Weaver}, {Y{\`e}che}, {Zou}, {Zhou}, {Abareshi},
  {Abbott}, {Abolfathi}, {Aguilera}, {Alam}, {Allen}, {Alvarez}, {Annis},
  {Ansarinejad}, {Aubert}, {Beechert}, {Bell}, {BenZvi}, {Beutler}, {Bielby},
  {Bolton}, {Brice{\~n}o}, {Buckley-Geer}, {Butler}, {Calamida}, {Carlberg},
  {Carter}, {Casas}, {Castander}, {Choi}, {Comparat}, {Cukanovaite}, {Delubac},
  {DeVries}, {Dey}, {Dhungana}, {Dickinson}, {Ding}, {Donaldson}, {Duan},
  {Duckworth}, {Eftekharzadeh}, {Eisenstein}, {Etourneau}, {Fagrelius},
  {Farihi}, {Fitzpatrick}, {Font-Ribera}, {Fulmer}, {G{\"a}nsicke},
  {Gaztanaga}, {George}, {Gerdes}, {Gontcho}, {Gorgoni}, {Green}, {Guy},
  {Harmer}, {Hernandez}, {Honscheid}, {Huang}, {James}, {Jannuzi}, {Jiang},
  {Joyce}, {Karcher}, {Karkar}, {Kehoe}, {Kneib}, {Kueter-Young}, {Lan},
  {Lauer}, {Le Guillou}, {Le Van Suu}, {Lee}, {Lesser}, {Perreault Levasseur},
  {Li}, {Mann}, {Marshall}, {Mart{\'\i}nez-V{\'a}zquez}, {Martini}, {du Mas des
  Bourboux}, {McManus}, {Meier}, {M{\'e}nard}, {Metcalfe},
  {Mu{\~n}oz-Guti{\'e}rrez}, {Najita}, {Napier}, {Narayan}, {Newman}, {Nie},
  {Nord}, {Norman}, {Olsen}, {Paat}, {Palanque-Delabrouille}, {Peng},
  {Poppett}, {Poremba}, {Prakash}, {Rabinowitz}, {Raichoor}, {Rezaie},
  {Robertson}, {Roe}, {Ross}, {Ross}, {Rudnick}, {Safonova}, {Saha},
  {S{\'a}nchez}, {Savary}, {Schweiker}, {Scott}, {Seo}, {Shan}, {Silva},
  {Slepian}, {Soto}, {Sprayberry}, {Staten}, {Stillman}, {Stupak}, {Summers},
  {Sien Tie}, {Tirado}, {Vargas-Maga{\~n}a}, {Vivas}, {Wechsler}, {Williams},
  {Yang}, {Yang}, {Yapici}, {Zaritsky}, {Zenteno}, {Zhang}, {Zhang}, {Zhou}, \&
  {Zhou}}]{2019AJ....157..168D}
{Dey}, A., {Schlegel}, D.~J., {Lang}, D., {et~al.} 2019,
  \bibinfo{title}{{Overview of the DESI Legacy Imaging Surveys},} \aj, 157,
  168, \dodoi{10.3847/1538-3881/ab089d}

\bibitem[{S.~A. {Dodd} {et~al.}(2025){Dodd}, {Huang}, {Davis}, \&
  {Ramirez-Ruiz}}]{Dodd2025}
{Dodd}, S.~A., {Huang}, X., {Davis}, S.~W., \& {Ramirez-Ruiz}, E. 2025,
  \bibinfo{title}{{Perturbing AGN Accretion Disks with Stars and Moderately
  Massive Black Holes: Implications for Changing-Look AGN and Quasi-Periodic
  Eruptions},} arXiv e-prints, arXiv:2506.19900,
  \dodoi{10.48550/arXiv.2506.19900}

\bibitem[{S.~A. {Dodd} {et~al.}(2021){Dodd}, {Law-Smith}, {Auchettl},
  {Ramirez-Ruiz}, \& {Foley}}]{2021_Dodd}
{Dodd}, S.~A., {Law-Smith}, J. A.~P., {Auchettl}, K., {Ramirez-Ruiz}, E., \&
  {Foley}, R.~J. 2021, \bibinfo{title}{{The Landscape of Galaxies Harboring
  Changing-look Active Galactic Nuclei in the Local Universe},} \apjl, 907,
  L21, \dodoi{10.3847/2041-8213/abd852}

\bibitem[{S.~A. {Dodd} {et~al.}(2023){Dodd}, {Nukala}, {Connor}, {Auchettl},
  {French}, {Law-Smith}, {Hammerstein}, \&
  {Ramirez-Ruiz}}]{2023ApJ...959L..19D}
{Dodd}, S.~A., {Nukala}, A., {Connor}, I., {et~al.} 2023,
  \bibinfo{title}{{Mid-infrared Outbursts in Nearby Galaxies: Nuclear
  Obscuration and Connections to Hidden Tidal Disruption Events and
  Changing-look Active Galactic Nuclei},} \apjl, 959, L19,
  \dodoi{10.3847/2041-8213/ad1112}

\bibitem[{A. {Dumont} {et~al.}(2025){Dumont}, {Neumayer}, {Seth}, {B{\"o}ker},
  {Eracleous}, {Goold}, {Greene}, {G{\"u}ltekin}, {Ho}, {Walsh}, \&
  {L{\"u}tzgendorf}}]{Dumont_2025_wicked}
{Dumont}, A., {Neumayer}, N., {Seth}, A.~C., {et~al.} 2025,
  \bibinfo{title}{{WIggle Corrector Kit for NIRSpEc Data: WICKED},} arXiv
  e-prints, arXiv:2503.09697, \dodoi{10.48550/arXiv.2503.09697}

\bibitem[{P. {Erwin} {et~al.}(2015){Erwin}, {Saglia}, {Fabricius}, {Thomas},
  {Nowak}, {Rusli}, {Bender}, {Vega Beltr{\'a}n}, \&
  {Beckman}}]{Erwin_2015_bulge}
{Erwin}, P., {Saglia}, R.~P., {Fabricius}, M., {et~al.} 2015,
  \bibinfo{title}{{Composite bulges: the coexistence of classical bulges and
  discy pseudo-bulges in S0 and spiral galaxies},} \mnras, 446, 4039,
  \dodoi{10.1093/mnras/stu2376}

\bibitem[{S. {Faris} {et~al.}(2024){Faris}, {Arcavi}, {Newsome}, {Farah},
  {Andrews}, {Howell}, \& {McCully}}]{Faris_2024_24tvdTNS}
{Faris}, S., {Arcavi}, I., {Newsome}, M., {et~al.} 2024,
  \bibinfo{title}{{StarDestroyers Transient Classification Report for
  2024-10-14},} Transient Name Server Classification Report, 2024-4005, 1

\bibitem[{D. {Foreman-Mackey} {et~al.}(2013){Foreman-Mackey}, {Hogg}, {Lang},
  \& {Goodman}}]{Foreman-Mackey_etal_2013_emcee}
{Foreman-Mackey}, D., {Hogg}, D.~W., {Lang}, D., \& {Goodman}, J. 2013,
  \bibinfo{title}{{emcee: The MCMC Hammer},} \pasp, 125, 306,
  \dodoi{10.1086/670067}

\bibitem[{D.~L. {Fried}(1966){Fried}}]{Fried_1966_Kolmogorov}
{Fried}, D.~L. 1966, \bibinfo{title}{{Optical Resolution Through a Randomly
  Inhomogeneous Medium for Very Long and Very Short Exposures},} Journal of the
  Optical Society of America (1917-1983), 56, 1372,
  \dodoi{10.1364/JOSA.56.001372}

\bibitem[{A.~S. {Fruchter} \& R.~N. {Hook}(2002){Fruchter} \&
  {Hook}}]{Fruchter_2002_Drizzle}
{Fruchter}, A.~S., \& {Hook}, R.~N. 2002, \bibinfo{title}{{Drizzle: A Method
  for the Linear Reconstruction of Undersampled Images},} \pasp, 114, 144,
  \dodoi{10.1086/338393}

\bibitem[{D.~A. {Gadotti}(2009){Gadotti}}]{Gadotti_2009_bulge}
{Gadotti}, D.~A. 2009, \bibinfo{title}{{Structural properties of pseudo-bulges,
  classical bulges and elliptical galaxies: a Sloan Digital Sky Survey
  perspective},} \mnras, 393, 1531, \dodoi{10.1111/j.1365-2966.2008.14257.x}

\bibitem[{S. {Gezari}(2021){Gezari}}]{Gezari_2021_tderev}
{Gezari}, S. 2021, \bibinfo{title}{{Tidal Disruption Events},} \araa, 59, 21,
  \dodoi{10.1146/annurev-astro-111720-030029}

\bibitem[{J.~E. {Greene} {et~al.}(2020){Greene}, {Strader}, \&
  {Ho}}]{Greene_2020_IMBHrev}
{Greene}, J.~E., {Strader}, J., \& {Ho}, L.~C. 2020,
  \bibinfo{title}{{Intermediate-Mass Black Holes},} \araa, 58, 257,
  \dodoi{10.1146/annurev-astro-032620-021835}

\bibitem[{J. {Guillochon} {et~al.}(2014){Guillochon}, {Manukian}, \&
  {Ramirez-Ruiz}}]{Guillochon_etal_2014}
{Guillochon}, J., {Manukian}, H., \& {Ramirez-Ruiz}, E. 2014,
  \bibinfo{title}{{PS1-10jh: The Disruption of a Main-sequence Star of
  Near-solar Composition},} \apj, 783, 23, \dodoi{10.1088/0004-637X/783/1/23}

\bibitem[{J. {Guillochon} {et~al.}(2018){Guillochon}, {Nicholl}, {Villar},
  {Mockler}, {Narayan}, {Mandel}, {Berger}, \&
  {Williams}}]{Guillochon_2018_mosfit}
{Guillochon}, J., {Nicholl}, M., {Villar}, V.~A., {et~al.} 2018,
  \bibinfo{title}{{MOSFiT: Modular Open Source Fitter for Transients},} \apjs,
  236, 6, \dodoi{10.3847/1538-4365/aab761}

\bibitem[{J. {Guillochon} \& E. {Ramirez-Ruiz}(2013){Guillochon} \&
  {Ramirez-Ruiz}}]{Guillochon_Ramirez_2013}
{Guillochon}, J., \& {Ramirez-Ruiz}, E. 2013, \bibinfo{title}{{Hydrodynamical
  Simulations to Determine the Feeding Rate of Black Holes by the Tidal
  Disruption of Stars: The Importance of the Impact Parameter and Stellar
  Structure},} \apj, 767, 25, \dodoi{10.1088/0004-637X/767/1/25}

\bibitem[{J. {Guillochon} \& E. {Ramirez-Ruiz}(2015){Guillochon} \&
  {Ramirez-Ruiz}}]{Guillochon2015}
{Guillochon}, J., \& {Ramirez-Ruiz}, E. 2015, \bibinfo{title}{{A Dark Year for
  Tidal Disruption Events},} \apj, 809, 166,
  \dodoi{10.1088/0004-637X/809/2/166}

\bibitem[{J.~E. {Gunn} {et~al.}(2006){Gunn}, {Siegmund}, {Mannery}, {Owen},
  {Hull}, {Leger}, {Carey}, {Knapp}, {York}, {Boroski}, {Kent}, {Lupton},
  {Rockosi}, {Evans}, {Waddell}, {Anderson}, {Annis}, {Barentine}, {Bartoszek},
  {Bastian}, {Bracker}, {Brewington}, {Briegel}, {Brinkmann}, {Brown}, {Carr},
  {Czarapata}, {Drennan}, {Dombeck}, {Federwitz}, {Gillespie}, {Gonzales},
  {Hansen}, {Harvanek}, {Hayes}, {Jordan}, {Kinney}, {Klaene}, {Kleinman},
  {Kron}, {Kresinski}, {Lee}, {Limmongkol}, {Lindenmeyer}, {Long}, {Loomis},
  {McGehee}, {Mantsch}, {Neilsen}, {Neswold}, {Newman}, {Nitta}, {Peoples},
  {Pier}, {Prieto}, {Prosapio}, {Rivetta}, {Schneider}, {Snedden}, \&
  {Wang}}]{Gunn_2006_SDSS}
{Gunn}, J.~E., {Siegmund}, W.~A., {Mannery}, E.~J., {et~al.} 2006,
  \bibinfo{title}{{The 2.5 m Telescope of the Sloan Digital Sky Survey},} \aj,
  131, 2332, \dodoi{10.1086/500975}

\bibitem[{M. {Guolo} \& A. {Mummery}(2025){Guolo} \&
  {Mummery}}]{Guolo_2025_disksize}
{Guolo}, M., \& {Mummery}, A. 2025, \bibinfo{title}{{The Size of Accretion
  Disks from Self-consistent X-Ray Spectra and UV/Optical/NIR Photometry
  Fitting: Applications to ASASSN{\textendash}14li and HLX{\textendash}1},}
  \apj, 978, 167, \dodoi{10.3847/1538-4357/ad990a}

\bibitem[{A. {Hajela} {et~al.}(2025){Hajela}, {Alexander}, {Margutti},
  {Chornock}, {Bietenholz}, {Christy}, {Stroh}, {Terreran}, {Saxton},
  {Komossa}, {Bright}, {Ramirez-Ruiz}, {Coppejans}, {Leung}, {Cendes},
  {Wiston}, {Laskar}, {Horesh}, {Schroeder}, {A.~J.}, {Wieringa}, {Velez},
  {Berger}, {Blanchard}, {Eftekhari}, {Gomez}, {Nicholl}, {Sears}, \&
  {Zauderer}}]{2025ApJ...983...29H}
{Hajela}, A., {Alexander}, K.~D., {Margutti}, R., {et~al.} 2025,
  \bibinfo{title}{{Eight Years of Light from ASASSN-15oi: Toward Understanding
  the Late-time Evolution of TDEs},} \apj, 983, 29,
  \dodoi{10.3847/1538-4357/adb620}

\bibitem[{E. {Hammerstein} {et~al.}(2023){Hammerstein}, {van Velzen}, {Gezari},
  {Cenko}, {Yao}, {Ward}, {Frederick}, {Villanueva}, {Somalwar}, {Graham},
  {Kulkarni}, {Stern}, {Andreoni}, {Bellm}, {Dekany}, {Dhawan}, {Drake},
  {Fremling}, {Gatkine}, {Groom}, {Ho}, {Kasliwal}, {Karambelkar}, {Kool},
  {Masci}, {Medford}, {Perley}, {Purdum}, {van Roestel}, {Sharma}, {Sollerman},
  {Taggart}, \& {Yan}}]{Hammerstein_2023_ZTFTDEs}
{Hammerstein}, E., {van Velzen}, S., {Gezari}, S., {et~al.} 2023,
  \bibinfo{title}{{The Final Season Reimagined: 30 Tidal Disruption Events from
  the ZTF-I Survey},} \apj, 942, 9, \dodoi{10.3847/1538-4357/aca283}

\bibitem[{C.~H. {Hannah} {et~al.}(2024){Hannah}, {Seth}, {Stone}, \& {van
  Velzen}}]{2024_Hannah}
{Hannah}, C.~H., {Seth}, A.~C., {Stone}, N.~C., \& {van Velzen}, S. 2024,
  \bibinfo{title}{{Counting the Unseen. I. Nuclear Density Scaling Relations
  for Nucleated Galaxies},} \aj, 168, 137, \dodoi{10.3847/1538-3881/ad630a}

\bibitem[{C.~H. {Hannah} {et~al.}(2025){Hannah}, {Stone}, {Seth}, \& {van
  Velzen}}]{2025_Hannah}
{Hannah}, C.~H., {Stone}, N.~C., {Seth}, A.~C., \& {van Velzen}, S. 2025,
  \bibinfo{title}{{Counting the Unseen. II. Tidal Disruption Event Rates in
  Nearby Galaxies with REPTiDE},} \apj, 988, 29,
  \dodoi{10.3847/1538-4357/addd1b}

\bibitem[{ {HI4PI Collaboration} {et~al.}(2016){HI4PI Collaboration}, {Ben
  Bekhti}, {Fl{\"o}er}, {Keller}, {Kerp}, {Lenz}, {Winkel}, {Bailin},
  {Calabretta}, {Dedes}, {Ford}, {Gibson}, {Haud}, {Janowiecki}, {Kalberla},
  {Lockman}, {McClure-Griffiths}, {Murphy}, {Nakanishi}, {Pisano}, \&
  {Staveley-Smith}}]{2016_HIPI4}
{HI4PI Collaboration}, {Ben Bekhti}, N., {Fl{\"o}er}, L., {et~al.} 2016,
  \bibinfo{title}{{HI4PI: A full-sky H I survey based on EBHIS and GASS},}
  \aap, 594, A116, \dodoi{10.1051/0004-6361/201629178}

\bibitem[{J.~G. {Hills}(1975){Hills}}]{Hills_1975}
{Hills}, J.~G. 1975, \bibinfo{title}{{Possible power source of Seyfert galaxies
  and QSOs},} \nat, 254, 295, \dodoi{10.1038/254295a0}

\bibitem[{P.~F. {Hopkins} {et~al.}(2010){Hopkins}, {Croton}, {Bundy},
  {Khochfar}, {van den Bosch}, {Somerville}, {Wetzel}, {Keres}, {Hernquist},
  {Stewart}, {Younger}, {Genel}, \& {Ma}}]{Hopkins_2010_galmerg}
{Hopkins}, P.~F., {Croton}, D., {Bundy}, K., {et~al.} 2010,
  \bibinfo{title}{{Mergers in {\ensuremath{\Lambda}}CDM: Uncertainties in
  Theoretical Predictions and Interpretations of the Merger Rate},} \apj, 724,
  915, \dodoi{10.1088/0004-637X/724/2/915}

\bibitem[{N. {Jiang} {et~al.}(2016){Jiang}, {Dou}, {Wang}, {Yang}, {Lyu}, \&
  {Zhou}}]{Jiang_2016_dustecho}
{Jiang}, N., {Dou}, L., {Wang}, T., {et~al.} 2016, \bibinfo{title}{{The WISE
  Detection of an Infrared Echo in Tidal Disruption Event ASASSN-14li},} \apjl,
  828, L14, \dodoi{10.3847/2041-8205/828/1/L14}

\bibitem[{N. {Jiang} {et~al.}(2021){Jiang}, {Wang}, {Hu}, {Sun}, {Dou}, \&
  {Xiao}}]{Jiang_2021_IRechoes}
{Jiang}, N., {Wang}, T., {Hu}, X., {et~al.} 2021, \bibinfo{title}{{Infrared
  Echoes of Optical Tidal Disruption Events: {\ensuremath{\sim}}1\%
  Dust-covering Factor or Less at Subparsec Scale},} \apj, 911, 31,
  \dodoi{10.3847/1538-4357/abe772}

\bibitem[{C.~C. {Jin} {et~al.}(2025){Jin}, {Li}, {Jiang}, {Dai}, {Cheng},
  {Zhu}, {Yang}, {Rau}, {Baldini}, {Wang}, {Zhou}, {Yuan}, {Zhang}, {Shu},
  {Shen}, {Wang}, {Wen}, {Wu}, {Wang}, {Thomsen}, {Zhang}, {Zhang}, {Coleiro},
  {Eyles-Ferris}, {Fang}, {Ho}, {Hu}, {Jin}, {Li}, {Liu}, {Liu}, {Liu}, {Liu},
  {Lu}, {Merloni}, {Qiao}, {Saxton}, {Soria}, {Wang}, {Xue}, {Yang}, {Zhang},
  {Zhang}, {Cai}, {Chen}, {Chen}, {Chen}, {Chen}, {Chen}, {Chen}, {Chen},
  {Cordier}, {Cui}, {Cui}, {Dai}, {Ding}, {Fan}, {Fan}, {Feng}, {Garcia},
  {Guan}, {Han}, {Hou}, {Hu}, {Huang}, {Huo}, {Jia}, {Jia}, {Jiang}, {Jin},
  {Kong}, {Kuulkers}, {Lei}, {Li}, {Li}, {Li}, {Li}, {Li}, {Li}, {Lian},
  {Ling}, {Liu}, {Liu}, {Liu}, {Liu}, {Liu}, {Lu}, {Luo}, {Ma}, {Mao}, {Mu},
  {Nandra}, {O'Brien}, {Pan}, {Pan}, {Qin}, {Rea}, {Sanders}, {Song}, {Sun},
  {Sun}, {Sun}, {Tan}, {Tang}, {Tao}, {Wang}, {Wang}, {Wang}, {Wang}, {Wang},
  {Wang}, {Wang}, {Wu}, {Wu}, {Xu}, {Xu}, {Xu}, {Xu}, {Xu}, {Xue}, {Xue},
  {Xue}, {Yan}, {Yang}, {Yang}, {Zhang}, {Zhang}, {Zhang}, {Zhang}, {Zhang},
  {Zhang}, {Zhang}, {Zhao}, {Zhao}, {Zhao}, {Zhao}, {Zheng}, {Zhu}, {Zhu},
  {Zhu}, \& {Zou}}]{Jin_2025_EP24}
{Jin}, C.~C., {Li}, D.~Y., {Jiang}, N., {et~al.} 2025, \bibinfo{title}{{An
  Intermediate-mass Black Hole Lurking in A Galactic Halo Caught Alive during
  Outburst},} arXiv e-prints, arXiv:2501.09580,
  \dodoi{10.48550/arXiv.2501.09580}

\bibitem[{R.~E. Kass \& A.~E. Raftery(1995)Kass \& Raftery}]{kass1995bayes}
Kass, R.~E., \& Raftery, A.~E. 1995, \bibinfo{title}{Bayes Factors,} Journal of
  the American Statistical Association, 90, 773,
  \dodoi{10.1080/01621459.1995.10476572}

\bibitem[{M. {Kesden}(2012){Kesden}}]{Kesden2012}
{Kesden}, M. 2012, \bibinfo{title}{{Black-hole spin dependence in the light
  curves of tidal disruption events},} \prd, 86, 064026,
  \dodoi{10.1103/PhysRevD.86.064026}

\bibitem[{S. {Komossa}(2012){Komossa}}]{Komossa_2012_recoilBHrev}
{Komossa}, S. 2012, \bibinfo{title}{{Recoiling Black Holes: Electromagnetic
  Signatures, Candidates, and Astrophysical Implications},} Advances in
  Astronomy, 2012, 364973, \dodoi{10.1155/2012/364973}

\bibitem[{S. {Komossa}(2015){Komossa}}]{Komossa_2015_TDErev}
{Komossa}, S. 2015, \bibinfo{title}{{Tidal disruption of stars by supermassive
  black holes: Status of observations},} Journal of High Energy Astrophysics,
  7, 148, \dodoi{10.1016/j.jheap.2015.04.006}

\bibitem[{S. {Komossa} {et~al.}(2003){Komossa}, {Burwitz}, {Hasinger},
  {Predehl}, {Kaastra}, \& {Ikebe}}]{2003_Komossa}
{Komossa}, S., {Burwitz}, V., {Hasinger}, G., {et~al.} 2003,
  \bibinfo{title}{{Discovery of a Binary Active Galactic Nucleus in the
  Ultraluminous Infrared Galaxy NGC 6240 Using Chandra},} \apjl, 582, L15,
  \dodoi{10.1086/346145}

\bibitem[{J. {Kormendy} \& L.~C. {Ho}(2013){Kormendy} \&
  {Ho}}]{Kormendy_2013_Msigma}
{Kormendy}, J., \& {Ho}, L.~C. 2013, \bibinfo{title}{{Coevolution (Or Not) of
  Supermassive Black Holes and Host Galaxies},} \araa, 51, 511,
  \dodoi{10.1146/annurev-astro-082708-101811}

\bibitem[{J. {Kormendy} \& R.~C. {Kennicutt}(2004){Kormendy} \&
  {Kennicutt}}]{Kormendy_2004_bulge}
{Kormendy}, J., \& {Kennicutt}, Jr., R.~C. 2004, \bibinfo{title}{{Secular
  Evolution and the Formation of Pseudobulges in Disk Galaxies},} \araa, 42,
  603, \dodoi{10.1146/annurev.astro.42.053102.134024}

\bibitem[{P. {Kroupa}(2001){Kroupa}}]{2001_Kroupa}
{Kroupa}, P. 2001, \bibinfo{title}{{On the variation of the initial mass
  function},} \mnras, 322, 231, \dodoi{10.1046/j.1365-8711.2001.04022.x}

\bibitem[{J. {Law-Smith} {et~al.}(2017){Law-Smith}, {MacLeod}, {Guillochon},
  {Macias}, \& {Ramirez-Ruiz}}]{Law-Smith2017}
{Law-Smith}, J., {MacLeod}, M., {Guillochon}, J., {Macias}, P., \&
  {Ramirez-Ruiz}, E. 2017, \bibinfo{title}{{Low-mass White Dwarfs with Hydrogen
  Envelopes as a Missing Link in the Tidal Disruption Menu},} \apj, 841, 132,
  \dodoi{10.3847/1538-4357/aa6ffb}

\bibitem[{L.-X. {Li} {et~al.}(2005){Li}, {Zimmerman}, {Narayan}, \&
  {McClintock}}]{Li_2005_MCD}
{Li}, L.-X., {Zimmerman}, E.~R., {Narayan}, R., \& {McClintock}, J.~E. 2005,
  \bibinfo{title}{{Multitemperature Blackbody Spectrum of a Thin Accretion Disk
  around a Kerr Black Hole: Model Computations and Comparison with
  Observations},} \apjs, 157, 335, \dodoi{10.1086/428089}

\bibitem[{D. {Lin} {et~al.}(2018){Lin}, {Strader}, {Carrasco}, {Page},
  {Romanowsky}, {Homan}, {Irwin}, {Remillard}, {Godet}, {Webb}, {Baumgardt},
  {Wijnands}, {Barret}, {Duc}, {Brodie}, \& {Gwyn}}]{Lin_2018_J215}
{Lin}, D., {Strader}, J., {Carrasco}, E.~R., {et~al.} 2018, \bibinfo{title}{{A
  luminous X-ray outburst from an intermediate-mass black hole in an off-centre
  star cluster},} Nature Astronomy, 2, 656, \dodoi{10.1038/s41550-018-0493-1}

\bibitem[{D. {Lin} {et~al.}(2020){Lin}, {Strader}, {Romanowsky}, {Irwin},
  {Godet}, {Barret}, {Webb}, {Homan}, \& {Remillard}}]{Lin_2020_J215}
{Lin}, D., {Strader}, J., {Romanowsky}, A.~J., {et~al.} 2020,
  \bibinfo{title}{{Multiwavelength Follow-up of the Hyperluminous
  Intermediate-mass Black Hole Candidate 3XMM J215022.4-055108},} \apjl, 892,
  L25, \dodoi{10.3847/2041-8213/ab745b}

\bibitem[{W. {Lu} \& C. {Bonnerot}(2020){Lu} \& {Bonnerot}}]{Lu_Bonnerot_2020}
{Lu}, W., \& {Bonnerot}, C. 2020, \bibinfo{title}{{Self-intersection of the
  fallback stream in tidal disruption events},} \mnras, 492, 686,
  \dodoi{10.1093/mnras/stz3405}

\bibitem[{M. {MacLeod} {et~al.}(2012){MacLeod}, {Guillochon}, \&
  {Ramirez-Ruiz}}]{2012_Macleod}
{MacLeod}, M., {Guillochon}, J., \& {Ramirez-Ruiz}, E. 2012,
  \bibinfo{title}{{The Tidal Disruption of Giant Stars and their Contribution
  to the Flaring Supermassive Black Hole Population},} \apj, 757, 134,
  \dodoi{10.1088/0004-637X/757/2/134}

\bibitem[{M. {MacLeod} {et~al.}(2016){MacLeod}, {Trenti}, \&
  {Ramirez-Ruiz}}]{2016_MacLeod}
{MacLeod}, M., {Trenti}, M., \& {Ramirez-Ruiz}, E. 2016, \bibinfo{title}{{The
  Close Stellar Companions to Intermediate-mass Black Holes},} \apj, 819, 70,
  \dodoi{10.3847/0004-637X/819/1/70}

\bibitem[{J. {Magorrian} \& S. {Tremaine}(1999){Magorrian} \&
  {Tremaine}}]{Magorrian_1999_TDE-NSC}
{Magorrian}, J., \& {Tremaine}, S. 1999, \bibinfo{title}{{Rates of tidal
  disruption of stars by massive central black holes},} \mnras, 309, 447,
  \dodoi{10.1046/j.1365-8711.1999.02853.x}

\bibitem[{N.~J. {McConnell} \& C.-P. {Ma}(2013){McConnell} \&
  {Ma}}]{McConnell_2013_Msigma}
{McConnell}, N.~J., \& {Ma}, C.-P. 2013, \bibinfo{title}{{Revisiting the
  Scaling Relations of Black Hole Masses and Host Galaxy Properties},} \apj,
  764, 184, \dodoi{10.1088/0004-637X/764/2/184}

\bibitem[{P. {Melchior} {et~al.}(2018){Melchior}, {Moolekamp}, {Jerdee},
  {Armstrong}, {Sun}, {Bosch}, \& {Lupton}}]{2018A&C....24..129M}
{Melchior}, P., {Moolekamp}, F., {Jerdee}, M., {et~al.} 2018,
  \bibinfo{title}{{SCARLET: Source separation in multi-band images by
  Constrained Matrix Factorization},} Astronomy and Computing, 24, 129,
  \dodoi{10.1016/j.ascom.2018.07.001}

\bibitem[{D. {Melchor} {et~al.}(2024){Melchor}, {Mockler}, {Naoz}, {Rose}, \&
  {Ramirez-Ruiz}}]{2024_Melchor}
{Melchor}, D., {Mockler}, B., {Naoz}, S., {Rose}, S.~C., \& {Ramirez-Ruiz}, E.
  2024, \bibinfo{title}{{Tidal Disruption Events from the Combined Effects of
  Two-body Relaxation and the Eccentric Kozai-Lidov Mechanism},} \apj, 960, 39,
  \dodoi{10.3847/1538-4357/acfee0}

\bibitem[{B. {Mockler} {et~al.}(2019){Mockler}, {Guillochon}, \&
  {Ramirez-Ruiz}}]{2019_Mockler}
{Mockler}, B., {Guillochon}, J., \& {Ramirez-Ruiz}, E. 2019,
  \bibinfo{title}{{Weighing Black Holes Using Tidal Disruption Events},} \apj,
  872, 151, \dodoi{10.3847/1538-4357/ab010f}

\bibitem[{B. {Mockler} {et~al.}(2023){Mockler}, {Melchor}, {Naoz}, \&
  {Ramirez-Ruiz}}]{2023_Mockler}
{Mockler}, B., {Melchor}, D., {Naoz}, S., \& {Ramirez-Ruiz}, E. 2023,
  \bibinfo{title}{{Uncovering Hidden Massive Black Hole Companions with Tidal
  Disruption Events},} \apj, 959, 18, \dodoi{10.3847/1538-4357/ad0234}

\bibitem[{A.~F.~J. {Moffat}(1969){Moffat}}]{Moffat_1969_psf}
{Moffat}, A.~F.~J. 1969, \bibinfo{title}{{A Theoretical Investigation of Focal
  Stellar Images in the Photographic Emulsion and Application to Photographic
  Photometry},} \aap, 3, 455

\bibitem[{P. {Morrissey} {et~al.}(2018){Morrissey}, {Matuszewski}, {Martin},
  {Neill}, {Epps}, {Fucik}, {Weber}, {Darvish}, {Adkins}, {Allen}, {Bartos},
  {Belicki}, {Cabak}, {Callahan}, {Cowley}, {Crabill}, {Deich}, {Delecroix},
  {Doppman}, {Hilyard}, {James}, {Kaye}, {Kokorowski}, {Kwok}, {Lanclos},
  {Milner}, {Moore}, {O'Sullivan}, {Parihar}, {Park}, {Phillips}, {Rizzi},
  {Rockosi}, {Rodriguez}, {Salaun}, {Seaman}, {Sheikh}, {Weiss}, \&
  {Zarzaca}}]{Morrissey_2018_KCWI}
{Morrissey}, P., {Matuszewski}, M., {Martin}, D.~C., {et~al.} 2018,
  \bibinfo{title}{{The Keck Cosmic Web Imager Integral Field Spectrograph},}
  \apj, 864, 93, \dodoi{10.3847/1538-4357/aad597}

\bibitem[{D.~C. {Morton}(1991){Morton}}]{Morton_1991_vac_air}
{Morton}, D.~C. 1991, \bibinfo{title}{{Atomic Data for Resonance Absorption
  Lines. I. Wavelengths Longward of the Lyman Limit},} \apjs, 77, 119,
  \dodoi{10.1086/191601}

\bibitem[{A. {Mummery} \& S. {van Velzen}(2025){Mummery} \& {van
  Velzen}}]{Mummery_2025_TDEdisktheory}
{Mummery}, A., \& {van Velzen}, S. 2025, \bibinfo{title}{{The optical,
  UV-plateau, and X-ray tidal disruption event luminosity functions reproduced
  from first principles},} \mnras, 541, 429, \dodoi{10.1093/mnras/staf938}

\bibitem[{A. {Mummery} {et~al.}(2024){Mummery}, {van Velzen}, {Nathan},
  {Ingram}, {Hammerstein}, {Fraser-Taliente}, \&
  {Balbus}}]{Mummery_2024_BH-TDE_scaling}
{Mummery}, A., {van Velzen}, S., {Nathan}, E., {et~al.} 2024,
  \bibinfo{title}{{Fundamental scaling relationships revealed in the optical
  light curves of tidal disruption events},} \mnras, 527, 2452,
  \dodoi{10.1093/mnras/stad3001}

\bibitem[{T. {Naab} \& A. {Burkert}(2003){Naab} \&
  {Burkert}}]{Naab_2003_galmerg}
{Naab}, T., \& {Burkert}, A. 2003, \bibinfo{title}{{Statistical Properties of
  Collisionless Equal- and Unequal-Mass Merger Remnants of Disk Galaxies},}
  \apj, 597, 893, \dodoi{10.1086/378581}

\bibitem[{T. {Naab} {et~al.}(2009){Naab}, {Johansson}, \&
  {Ostriker}}]{Naab_2009_minormerg}
{Naab}, T., {Johansson}, P.~H., \& {Ostriker}, J.~P. 2009,
  \bibinfo{title}{{Minor Mergers and the Size Evolution of Elliptical
  Galaxies},} \apjl, 699, L178, \dodoi{10.1088/0004-637X/699/2/L178}

\bibitem[{S. {Naoz} {et~al.}(2020){Naoz}, {Will}, {Ramirez-Ruiz}, {Hees},
  {Ghez}, \& {Do}}]{2020_Naoz}
{Naoz}, S., {Will}, C.~M., {Ramirez-Ruiz}, E., {et~al.} 2020,
  \bibinfo{title}{{A Hidden Friend for the Galactic Center Black Hole, Sgr
  A*},} \apjl, 888, L8, \dodoi{10.3847/2041-8213/ab5e3b}

\bibitem[{ {Nasa High Energy Astrophysics Science Archive Research Center
  (Heasarc)}(2014){Nasa High Energy Astrophysics Science Archive Research
  Center (Heasarc)}}]{HEASOFT}
{Nasa High Energy Astrophysics Science Archive Research Center (Heasarc)}.
  2014, \bibinfo{title}{{HEAsoft: Unified Release of FTOOLS and XANADU},},
  Astrophysics Source Code Library \doeprint{1408.004}

\bibitem[{D. {Neill} {et~al.}(2023){Neill}, {Matuszewski}, {Martin},
  {Brodheim}, \& {Rizzi}}]{Neill_2023_KCWIDRP}
{Neill}, D., {Matuszewski}, M., {Martin}, C., {Brodheim}, M., \& {Rizzi}, L.
  2023, \bibinfo{title}{{KCWI\_DRP: Keck Cosmic Web Imager Data Reduction
  Pipeline in Python},}, Astrophysics Source Code Library, record ascl:2301.019
  \doeprint{2301.019}

\bibitem[{H. {Netzer}(2015){Netzer}}]{Netzer_2015_AGNrev}
{Netzer}, H. 2015, \bibinfo{title}{{Revisiting the Unified Model of Active
  Galactic Nuclei},} \araa, 53, 365,
  \dodoi{10.1146/annurev-astro-082214-122302}

\bibitem[{N. {Neumayer} {et~al.}(2020){Neumayer}, {Seth}, \&
  {B{\"o}ker}}]{Neumayer_2020_NSCrev}
{Neumayer}, N., {Seth}, A., \& {B{\"o}ker}, T. 2020, \bibinfo{title}{{Nuclear
  star clusters},} \aapr, 28, 4, \dodoi{10.1007/s00159-020-00125-0}

\bibitem[{M.~A. {Norris} {et~al.}(2014){Norris}, {Kannappan}, {Forbes},
  {Romanowsky}, {Brodie}, {Faifer}, {Huxor}, {Maraston}, {Moffett}, {Penny},
  {Pota}, {Smith-Castelli}, {Strader}, {Bradley}, {Eckert}, {Fohring},
  {McBride}, {Stark}, \& {Vaduvescu}}]{Norris_2014_UCD}
{Norris}, M.~A., {Kannappan}, S.~J., {Forbes}, D.~A., {et~al.} 2014,
  \bibinfo{title}{{The AIMSS Project - I. Bridging the star cluster-galaxy
  divide$^{★}${\textdagger}{\textdaggerdbl}{\textsection}{\textparagraph}},}
  \mnras, 443, 1151, \dodoi{10.1093/mnras/stu1186}

\bibitem[{I.~D. {Novikov} \& K.~S. {Thorne}(1973){Novikov} \&
  {Thorne}}]{Novikov_1973_accretion}
{Novikov}, I.~D., \& {Thorne}, K.~S. 1973, in Black Holes (Les Astres Occlus),
  ed. C.~{Dewitt} \& B.~S. {Dewitt}, 343--450

\bibitem[{F. {Onori} {et~al.}(2019){Onori}, {Cannizzaro}, {Jonker}, {Fraser},
  {Kostrzewa-Rutkowska}, {Martin-Carrillo}, {Benetti}, {Elias-Rosa},
  {Gromadzki}, {Harmanen}, {Mattila}, {Strizinger}, {Terreran}, \&
  {Wevers}}]{Onori_2019_bowen}
{Onori}, F., {Cannizzaro}, G., {Jonker}, P.~G., {et~al.} 2019,
  \bibinfo{title}{{Optical follow-up of the tidal disruption event iPTF16fnl:
  new insights from X-shooter observations},} \mnras, 489, 1463,
  \dodoi{10.1093/mnras/stz2053}

\bibitem[{F. {Onori} {et~al.}(2025){Onori}, {Nicholl}, {Ramsden}, {McGee},
  {Roy}, {Li}, {Arcavi}, {Anderson}, {Brocato}, {Bronikowski}, {Cenko},
  {Chambers}, {Chen}, {Clark}, {Concepcion}, {Farah}, {Flammini},
  {Gonz{\'a}lez-Gait{\'a}n}, {Gromadzki}, {Guti{\'e}rrez}, {Hammerstein},
  {Hinds}, {Inserra}, {Kankare}, {Kumar}, {Makrygianni}, {Mattila},
  {Matilainen}, {M{\"u}ller-Bravo}, {Petrushevska}, {Pignata}, {Piranomonte},
  {Reynolds}, {Stein}, {Wang}, {Wevers}, {Yao}, \& {Young}}]{2025_Onori}
{Onori}, F., {Nicholl}, M., {Ramsden}, P., {et~al.} 2025, \bibinfo{title}{{The
  case of AT2022wtn: a tidal disruption event in an interacting galaxy},}
  \mnras, 540, 498, \dodoi{10.1093/mnras/staf746}

\bibitem[{D.~N. {Page} \& K.~S. {Thorne}(1974){Page} \&
  {Thorne}}]{Page_1974_fr}
{Page}, D.~N., \& {Thorne}, K.~S. 1974, \bibinfo{title}{{Disk-Accretion onto a
  Black Hole. Time-Averaged Structure of Accretion Disk},} \apj, 191, 499,
  \dodoi{10.1086/152990}

\bibitem[{M. {Perrin} {et~al.}(2025){Perrin}, {Long}, {Osborne}, {Geda},
  {Sappington}, {Mel{\'e}ndez}, {Lajoie}, {Leisenring}, {Zimmerman}, {Brooks},
  {Otor}, {Kulp}, {Chambers}, \& {Jurling}}]{Perrin_2025_STPSF}
{Perrin}, M., {Long}, J., {Osborne}, S., {et~al.} 2025,
  \bibinfo{title}{{STPSF},}, 2.1.0 Zenodo, \dodoi{10.5281/zenodo.15747364}

\bibitem[{J. {Pfeffer} \& H. {Baumgardt}(2013){Pfeffer} \&
  {Baumgardt}}]{Pfeffer_2013_UCD}
{Pfeffer}, J., \& {Baumgardt}, H. 2013, \bibinfo{title}{{Ultra-compact dwarf
  galaxy formation by tidal stripping of nucleated dwarf galaxies},} \mnras,
  433, 1997, \dodoi{10.1093/mnras/stt867}

\bibitem[{H. {Pfister} {et~al.}(2020){Pfister}, {Volonteri}, {Dai}, \&
  {Colpi}}]{2020_Pfister}
{Pfister}, H., {Volonteri}, M., {Dai}, J.~L., \& {Colpi}, M. 2020,
  \bibinfo{title}{{Enhancement of the tidal disruption event rate in galaxies
  with a nuclear star cluster: from dwarfs to ellipticals},} \mnras, 497, 2276,
  \dodoi{10.1093/mnras/staa1962}

\bibitem[{M. {Polkas} {et~al.}(2024){Polkas}, {Bonoli}, {Bortolas},
  {Izquierdo-Villalba}, {Sesana}, {Broggi}, {Hoyer}, \&
  {Spinoso}}]{2024_Polkas}
{Polkas}, M., {Bonoli}, S., {Bortolas}, E., {et~al.} 2024,
  \bibinfo{title}{{Demographics of tidal disruption events with L-Galaxies: I.
  Volumetric TDE rates and the abundance of nuclear star clusters},} \aap, 689,
  A204, \dodoi{10.1051/0004-6361/202449470}

\bibitem[{E. {Ramirez-Ruiz} \& S. {Rosswog}(2009){Ramirez-Ruiz} \&
  {Rosswog}}]{2009_Ramirez-Ruiz}
{Ramirez-Ruiz}, E., \& {Rosswog}, S. 2009, \bibinfo{title}{{The Star Ingesting
  Luminosity of Intermediate-Mass Black Holes in Globular Clusters},} \apjl,
  697, L77, \dodoi{10.1088/0004-637X/697/2/L77}

\bibitem[{M.~J. {Rees}(1988){Rees}}]{Rees_1988}
{Rees}, M.~J. 1988, \bibinfo{title}{{Tidal disruption of stars by black holes
  of {}10$^{6}$-{}10$^{8}$ solar masses in nearby galaxies},} \nat, 333, 523,
  \dodoi{10.1038/333523a0}

\bibitem[{A. {Ricarte} {et~al.}(2021{\natexlab{a}}){Ricarte}, {Tremmel},
  {Natarajan}, \& {Quinn}}]{Ricarte_2021_wandBHA}
{Ricarte}, A., {Tremmel}, M., {Natarajan}, P., \& {Quinn}, T.
  2021{\natexlab{a}}, \bibinfo{title}{{Unveiling the Population of Wandering
  Black Holes via Electromagnetic Signatures},} \apjl, 916, L18,
  \dodoi{10.3847/2041-8213/ac1170}

\bibitem[{A. {Ricarte} {et~al.}(2021{\natexlab{b}}){Ricarte}, {Tremmel},
  {Natarajan}, {Zimmer}, \& {Quinn}}]{Ricarte_2021_wandBHB}
{Ricarte}, A., {Tremmel}, M., {Natarajan}, P., {Zimmer}, C., \& {Quinn}, T.
  2021{\natexlab{b}}, \bibinfo{title}{{Origins and demographics of wandering
  black holes},} \mnras, 503, 6098, \dodoi{10.1093/mnras/stab866}

\bibitem[{C. {Rodriguez} {et~al.}(2006){Rodriguez}, {Taylor}, {Zavala}, {Peck},
  {Pollack}, \& {Romani}}]{2006_Rodriguez}
{Rodriguez}, C., {Taylor}, G.~B., {Zavala}, R.~T., {et~al.} 2006,
  \bibinfo{title}{{A Compact Supermassive Binary Black Hole System},} \apj,
  646, 49, \dodoi{10.1086/504825}

\bibitem[{N. {Roth} {et~al.}(2016){Roth}, {Kasen}, {Guillochon}, \&
  {Ramirez-Ruiz}}]{Roth_etal_2016}
{Roth}, N., {Kasen}, D., {Guillochon}, J., \& {Ramirez-Ruiz}, E. 2016,
  \bibinfo{title}{{The X-Ray through Optical Fluxes and Line Strengths of Tidal
  Disruption Events},} \apj, 827, 3, \dodoi{10.3847/0004-637X/827/1/3}

\bibitem[{M. {Rozner} \& E. {Ramirez-Ruiz}(2025){Rozner} \&
  {Ramirez-Ruiz}}]{Rozner2025}
{Rozner}, M., \& {Ramirez-Ruiz}, E. 2025, \bibinfo{title}{{Stellar
  Distributions around Supermassive Black Holes in Gas-rich Nuclear Star
  Clusters},} \apjl, 988, L21, \dodoi{10.3847/2041-8213/adeca7}

\bibitem[{E.~F. {Schlafly} \& D.~P. {Finkbeiner}(2011){Schlafly} \&
  {Finkbeiner}}]{Schlafly_etal_2011}
{Schlafly}, E.~F., \& {Finkbeiner}, D.~P. 2011, \bibinfo{title}{{Measuring
  Reddening with Sloan Digital Sky Survey Stellar Spectra and Recalibrating
  SFD},} ApJ, 737, 103, \dodoi{10.1088/0004-637X/737/2/103}

\bibitem[{G. {Schwarz}(1978){Schwarz}}]{Schwarz_1978_BIC}
{Schwarz}, G. 1978, \bibinfo{title}{{Estimating the Dimension of a Model},}
  Annals of Statistics, 6, 461

\bibitem[{N. {Scott} \& A.~W. {Graham}(2013){Scott} \&
  {Graham}}]{Scott_2013_NSC}
{Scott}, N., \& {Graham}, A.~W. 2013, \bibinfo{title}{{Updated Mass Scaling
  Relations for Nuclear Star Clusters and a Comparison to Supermassive Black
  Holes},} \apj, 763, 76, \dodoi{10.1088/0004-637X/763/2/76}

\bibitem[{I. {Sfaradi} {et~al.}(2025){Sfaradi}, {Margutti}, {Chornock},
  {Alexander}, {Metzger}, {Beniamini}, {Barniol Duran}, {Yao}, {Horesh},
  {Farah}, {Berger}, {Nayana A.}, {Cendes}, {Eftekhari}, {Fender}, {Franz},
  {Green}, {Hammerstein}, {Lu}, {Wiston}, {Bernstein}, {Bright}, {Christy},
  {Cruz}, {DeBoer}, {Golay}, {Goodwin}, {Gurwell}, {Keating}, {Laskar},
  {Miller-Jones}, {Pollak}, {Rao}, {Siemion}, {Sheikh}, {Shoval}, \& {van
  Velzen}}]{Sfaradi_2025_24tvdradio}
{Sfaradi}, I., {Margutti}, R., {Chornock}, R., {et~al.} 2025,
  \bibinfo{title}{{The First Radio-Bright Off-Nuclear TDE 2024tvd Reveals the
  Fastest-Evolving Double-Peaked Radio Emission},} arXiv e-prints,
  arXiv:2508.03807, \dodoi{10.48550/arXiv.2508.03807}

\bibitem[{N.~I. {Shakura} \& R.~A. {Sunyaev}(1973){Shakura} \&
  {Sunyaev}}]{Shakura_1973_accretion}
{Shakura}, N.~I., \& {Sunyaev}, R.~A. 1973, \bibinfo{title}{{Black holes in
  binary systems. Observational appearance.},} \aap, 24, 337

\bibitem[{H. {Shiokawa} {et~al.}(2015){Shiokawa}, {Krolik}, {Cheng}, {Piran},
  \& {Noble}}]{Shiokawa_etal_2015}
{Shiokawa}, H., {Krolik}, J.~H., {Cheng}, R.~M., {Piran}, T., \& {Noble}, S.~C.
  2015, \bibinfo{title}{{General Relativistic Hydrodynamic Simulation of
  Accretion Flow from a Stellar Tidal Disruption},} \apj, 804, 85,
  \dodoi{10.1088/0004-637X/804/2/85}

\bibitem[{A. {Sillanpaa} {et~al.}(1988){Sillanpaa}, {Haarala}, {Valtonen},
  {Sundelius}, \& {Byrd}}]{1988_Sillanpaa}
{Sillanpaa}, A., {Haarala}, S., {Valtonen}, M.~J., {Sundelius}, B., \& {Byrd},
  G.~G. 1988, \bibinfo{title}{{OJ 287: Binary Pair of Supermassive Black
  Holes},} \apj, 325, 628, \dodoi{10.1086/166033}

\bibitem[{L. {Simard} {et~al.}(2011){Simard}, {Mendel}, {Patton}, {Ellison}, \&
  {McConnachie}}]{Simard_2011_galcat}
{Simard}, L., {Mendel}, J.~T., {Patton}, D.~R., {Ellison}, S.~L., \&
  {McConnachie}, A.~W. 2011, \bibinfo{title}{{A Catalog of Bulge+disk
  Decompositions and Updated Photometry for 1.12 Million Galaxies in the Sloan
  Digital Sky Survey},} \apjs, 196, 11, \dodoi{10.1088/0067-0049/196/1/11}

\bibitem[{A. {S{\k{a}}dowski} {et~al.}(2014){S{\k{a}}dowski}, {Narayan},
  {McKinney}, \& {Tchekhovskoy}}]{Sadowski_2014_disksim}
{S{\k{a}}dowski}, A., {Narayan}, R., {McKinney}, J.~C., \& {Tchekhovskoy}, A.
  2014, \bibinfo{title}{{Numerical simulations of super-critical black hole
  accretion flows in general relativity},} \mnras, 439, 503,
  \dodoi{10.1093/mnras/stt2479}

\bibitem[{N. {Stone} {et~al.}(2013){Stone}, {Sari}, \&
  {Loeb}}]{stone13_frozen_in}
{Stone}, N., {Sari}, R., \& {Loeb}, A. 2013, \bibinfo{title}{{Consequences of
  strong compression in tidal disruption events},} \mnras, 435, 1809,
  \dodoi{10.1093/mnras/stt1270}

\bibitem[{N.~C. {Stone} \& B.~D. {Metzger}(2016){Stone} \&
  {Metzger}}]{Stone_2016_TDE_MBH}
{Stone}, N.~C., \& {Metzger}, B.~D. 2016, \bibinfo{title}{{Rates of stellar
  tidal disruption as probes of the supermassive black hole mass function},}
  \mnras, 455, 859, \dodoi{10.1093/mnras/stv2281}

\bibitem[{L.~E. {Strubbe} \& E. {Quataert}(2009){Strubbe} \&
  {Quataert}}]{Strubbe_etal_2009}
{Strubbe}, L.~E., \& {Quataert}, E. 2009, \bibinfo{title}{{Optical flares from
  the tidal disruption of stars by massive black holes},} \mnras, 400, 2070,
  \dodoi{10.1111/j.1365-2966.2009.15599.x}

\bibitem[{L.~L. {Thomsen} {et~al.}(2022){Thomsen}, {Kwan}, {Dai}, {Wu}, {Roth},
  \& {Ramirez-Ruiz}}]{2022_Thomsen}
{Thomsen}, L.~L., {Kwan}, T.~M., {Dai}, L., {et~al.} 2022,
  \bibinfo{title}{{Dynamical Unification of Tidal Disruption Events},} \apjl,
  937, L28, \dodoi{10.3847/2041-8213/ac911f}

\bibitem[{M. {Tremmel} {et~al.}(2018){Tremmel}, {Governato}, {Volonteri},
  {Pontzen}, \& {Quinn}}]{Tremmel_2018_wandMBH}
{Tremmel}, M., {Governato}, F., {Volonteri}, M., {Pontzen}, A., \& {Quinn},
  T.~R. 2018, \bibinfo{title}{{Wandering Supermassive Black Holes in
  Milky-Way-mass Halos},} \apjl, 857, L22, \dodoi{10.3847/2041-8213/aabc0a}

\bibitem[{S. {van Velzen} {et~al.}(2016){van Velzen}, {Mendez}, {Krolik}, \&
  {Gorjian}}]{vanvelzen_2016_TDedust}
{van Velzen}, S., {Mendez}, A.~J., {Krolik}, J.~H., \& {Gorjian}, V. 2016,
  \bibinfo{title}{{Discovery of Transient Infrared Emission from Dust Heated by
  Stellar Tidal Disruption Flares},} \apj, 829, 19,
  \dodoi{10.3847/0004-637X/829/1/19}

\bibitem[{S. {van Velzen} {et~al.}(2021){van Velzen}, {Pasham}, {Komossa},
  {Yan}, \& {Kara}}]{vanVelzen_2021_dust}
{van Velzen}, S., {Pasham}, D.~R., {Komossa}, S., {Yan}, L., \& {Kara}, E.~A.
  2021, \bibinfo{title}{{Reverberation in Tidal Disruption Events: Dust Echoes,
  Coronal Emission Lines, Multi-wavelength Cross-correlations, and QPOs},}
  \ssr, 217, 63, \dodoi{10.1007/s11214-021-00835-6}

\bibitem[{S. {van Velzen} {et~al.}(2019){van Velzen}, {Stone}, {Metzger},
  {Gezari}, {Brown}, \& {Fruchter}}]{vanVelzen_2019_UVTDEdisk}
{van Velzen}, S., {Stone}, N.~C., {Metzger}, B.~D., {et~al.} 2019,
  \bibinfo{title}{{Late-time UV Observations of Tidal Disruption Flares Reveal
  Unobscured, Compact Accretion Disks},} \apj, 878, 82,
  \dodoi{10.3847/1538-4357/ab1844}

\bibitem[{S. van Velzen {et~al.}(2021)van Velzen, {Gezari}, {Hammerstein},
  {Roth}, {Frederick}, {Ward}, {Hung}, {Cenko}, {Stein}, {Perley}, {Taggart},
  {Foley}, {Sollerman}, {Blagorodnova}, {Andreoni}, {Bellm}, {Brinnel}, {De},
  {Dekany}, {Feeney}, {Fremling}, {Giomi}, {Golkhou}, {Graham}, {Ho},
  {Kasliwal}, {Kilpatrick}, {Kulkarni}, {Kupfer}, {Laher}, {Mahabal}, {Masci},
  {Miller}, {Nordin}, {Riddle}, {Rusholme}, {van Santen}, {Sharma}, {Shupe}, \&
  {Soumagnac}}]{vanvelzen21_TDE_tdes}
van Velzen, S., {Gezari}, S., {Hammerstein}, E., {et~al.} 2021,
  \bibinfo{title}{{Seventeen Tidal Disruption Events from the First Half of ZTF
  Survey Observations: Entering a New Era of Population Studies},} \apj, 908,
  4, \dodoi{10.3847/1538-4357/abc258}

\bibitem[{A. {Vazdekis} {et~al.}(2016){Vazdekis}, {Koleva}, {Ricciardelli},
  {R{\"o}ck}, \& {Falc{\'o}n-Barroso}}]{Vazdekis_2016_emiles}
{Vazdekis}, A., {Koleva}, M., {Ricciardelli}, E., {R{\"o}ck}, B., \&
  {Falc{\'o}n-Barroso}, J. 2016, \bibinfo{title}{{UV-extended E-MILES stellar
  population models: young components in massive early-type galaxies},} \mnras,
  463, 3409, \dodoi{10.1093/mnras/stw2231}

\bibitem[{M. {Volonteri} \& R. {Perna}(2005){Volonteri} \&
  {Perna}}]{Volonteri_Perna_2005_ejectMBH}
{Volonteri}, M., \& {Perna}, R. 2005, \bibinfo{title}{{Dynamical evolution of
  intermediate-mass black holes and their observable signatures in the nearby
  Universe},} \mnras, 358, 913, \dodoi{10.1111/j.1365-2966.2005.08832.x}

\bibitem[{J. {Wang} \& D. {Merritt}(2004){Wang} \&
  {Merritt}}]{Wang_merritt_2004_TDE-NSC}
{Wang}, J., \& {Merritt}, D. 2004, \bibinfo{title}{{Revised Rates of Stellar
  Disruption in Galactic Nuclei},} \apj, 600, 149, \dodoi{10.1086/379767}

\bibitem[{Y. {Yao} {et~al.}(2023){Yao}, {Ravi}, {Gezari}, {van Velzen}, {Lu},
  {Schulze}, {Somalwar}, {Kulkarni}, {Hammerstein}, {Nicholl}, {Graham},
  {Perley}, {Cenko}, {Stein}, {Ricarte}, {Chadayammuri}, {Quataert}, {Bellm},
  {Bloom}, {Dekany}, {Drake}, {Groom}, {Mahabal}, {Prince}, {Riddle},
  {Rusholme}, {Sharma}, {Sollerman}, \& {Yan}}]{Yao_Yuhan_2023_TDEs}
{Yao}, Y., {Ravi}, V., {Gezari}, S., {et~al.} 2023, \bibinfo{title}{{Tidal
  Disruption Event Demographics with the Zwicky Transient Facility: Volumetric
  Rates, Luminosity Function, and Implications for the Local Black Hole Mass
  Function},} \apjl, 955, L6, \dodoi{10.3847/2041-8213/acf216}

\bibitem[{Y. {Yao} {et~al.}(2025){Yao}, {Chornock}, {Ward}, {Hammerstein},
  {Sfaradi}, {Margutti}, {Kelley}, {Lu}, {Liu}, {Wise}, {Sollerman},
  {Alexander}, {Bellm}, {Drake}, {Fremling}, {Gilfanov}, {Graham}, {Groom},
  {Hinds}, {Kulkarni}, {Miller}, {Miller-Jones}, {Nicholl}, {Perley}, {Purdum},
  {Ravi}, {Rich}, {Rehemtulla}, {Riddle}, {Smith}, {Stein}, {Sunyaev}, {van
  Velzen}, \& {Wold}}]{Yao_Yuhan_2025_24tvd}
{Yao}, Y., {Chornock}, R., {Ward}, C., {et~al.} 2025, \bibinfo{title}{{A
  Massive Black Hole 0.8 kpc from the Host Nucleus Revealed by the Offset Tidal
  Disruption Event AT2024tvd},} \apjl, 985, L48,
  \dodoi{10.3847/2041-8213/add7de}

\bibitem[{E.~R. {Zimmerman} {et~al.}(2005){Zimmerman}, {Narayan}, {McClintock},
  \& {Miller}}]{Zimmerman_2005_ezdiskbb}
{Zimmerman}, E.~R., {Narayan}, R., {McClintock}, J.~E., \& {Miller}, J.~M.
  2005, \bibinfo{title}{{Multitemperature Blackbody Spectra of Thin Accretion
  Disks with and without a Zero-Torque Inner Boundary Condition},} \apj, 618,
  832, \dodoi{10.1086/426071}

\end{thebibliography}
\bibliographystyle{aasjournalv7}

\end{document}